\documentclass[10pt, conference, letterpaper]{IEEEtran}
\IEEEoverridecommandlockouts
\usepackage{subcaption}
\usepackage{cite}
\usepackage{amsmath,amssymb,amsfonts}
\usepackage{algorithm}
\usepackage{graphicx}
\usepackage{textcomp}
\usepackage{booktabs}
\usepackage{pifont}
\usepackage{xcolor}
\usepackage{enumitem}
\usepackage{authblk} 
\usepackage{url}

\usepackage{authblk}
\def\BibTeX{{\rm B\kern-.05em{\sc i\kern-.025em b}\kern-.08em
    T\kern-.1667em\lower.7ex\hbox{E}\kern-.125emX}}
\begin{document}

\title{\Large \bf CD-Raft: Reducing the Latency of Distributed \\ Consensus in Cross-Domain Sites
\thanks{This work is supported by the National Key R\&D Program of China under Grant 2022YFB4501703, the Jiangxi Provincial Career-Early Young Scientists and Technologists Cultivation Project under Grant 20252BEJ730003, and the Jiangxi Provincial Natural Science Foundation under Grant 20252BAC200615. Corresponding author: *Zichen Xu (xuz@ncu.edu.cn).}
}

\author[ ]{Yangyang Wang$^{\dagger}$$^{\ddagger}$}
\author[ ]{Ziqian Cheng$^{\dagger}$$^{\ddagger}$}
\author[ ]{Yucong Dong$^{\dagger}$$^{\ddagger}$}
\author[*]{Zichen Xu$^{\dagger}$$^{\ddagger}$}
\affil[ ]{$^{\dagger}$School of Artificial Intelligence, Nanchang University, China}
\affil[ ]{$^{\ddagger}$School of Mathematics and Computer Sciences, Nanchang University, China}

\maketitle

\vspace{-0.5cm}
\begin{abstract}
Today's massive AI computation loads push heavy data synchronization across sites, i.e., nodes in data centers. 
Any reduction in such consensus latency can significantly improve the overall performance of desired systems.
This consensus challenge explosively peaks at cross-domain sites. 
In this paper, we proposed CD-Raft to address the cross-domain latency challenge, an optimized Raft protocol for strong consistency in cross-domain sites.
CD-Raft can significantly reduce consensus latency by optimizing cross-domain round-trip time (RTT) for reads and writes, as well as carefully positioning the leader node.
We verified the correctness of CD-Raft in a formal specification using the TLA+ specification, guaranteeing the strong consistency across sites. 
We have prototyped CD-Raft and evaluated it using the YCSB benchmark. 
Empirical results show that  compared to the classic Raft, CD-Raft reduces the average latency by 32.90\% and (99th percentile) tail latency by 49.24\% for renown traces across multiple sites.
\end{abstract}



\section{Introduction}

Large-scale AI computational workloads, especially those involving data or task parallelism, often require frequent synchronization of state and data across different devices to sites, in order to ensure consistency and accuracy.
This strong consistency 
actions fork unprecedented overhead in the overall computation performance.
What is worse, such workloads usually invoke more devices than one single site holds, thus requiring geographically distributed consistency across different sites, leading to frequent reads/writes at least in the Second-Level Domain (SLD).
Such cross-domain data synchronization puts a simple yet challenging requirement for today's consistency protocol: a light-weight consistency framework under heavy tail network traffic in distance.
For example, a single data transmission for a 7-billion-parameter (7B) model can be as large as 28GB~\cite{pretrain}. 

Therefore, the demand for massive data synchronization,
combined with high cross-domain communication latency,
renders the latency of the underlying consensus protocol the
decisive bottleneck for system performance.
For leader-based distributed consensus protocols, a write request typically involves two cross-domain RTTs (long RTTs): the first cross-domain RTT occurs between the client and the leader (when the client and the leader are in different domains) for the sending and acknowledgment of the request.
The second cross-domain RTT is performed between the leader and nodes in other domains to achieve data consistency confirmation.
In this case, the cross-domain latency in leader-based distributed consensus protocols significantly impacts the write request response time in distributed systems.
Therefore, optimizing this communication process is essential to improve system performance and reduce latency.
In addition, the choice of the leader's position directly affects communication efficiency, so optimizing the leader's position strategy is important.

To address the aforementioned challenge, we propose CD-Raft, an optimized Raft protocol suitable for cross-domain sites.
CD-Raft introduces two leader roles, the \textit{Domain Leader} and the \textit{Global Leader}.
In CD-Raft, each domain’s consensus operations are managed by the \textit{Domain Leader}, while the \textit{Global Leader} coordinates consensus operations across all \textit{Domain Leaders}.
Additionally, the \textit{Domain Leader} can act as a proxy for the \textit{Global Leader} in responding to the client.
CD-Raft is specifically optimized for cross-domain sites and incorporates two key strategies—\textit{Fast Return} and \textit{Optimal Global Leader Position}—designed to accelerate cross-domain operations.
Through these strategies, CD-Raft reduces the number of cross-domain RTTs for client requests, minimizing the total global latency while ensuring cross-domain disaster tolerance.

We have implemented a CD-Raft-based prototype system and rigorously evaluated its performance emulating the YCSB benchmark\cite{a5}.
The experimental results indicate that CD-Raft substantially enhances system performance, reducing the read and write latency across various workloads.
For example, compared to Raft, CD-Raft reduces the average latency by 39.81\% and tail latency by 49.44\% in a write-only workload.
In a typical read/write-balanced workload, CD-Raft reduces the average latency by 32.90\% and the tail latency by 49.24\%.

The main contributions of this paper are as follows:
\begin{itemize}[leftmargin=*]
    \item {We identify the performance bottlenecks of leader-based distributed consensus protocols in cross-domain environments. To ensure cross-domain disaster tolerance, leader-based distributed consensus protocols face two cross-domain RTTs during communication (when the client and the leader are in different domains) and data synchronization.}
    \item{We propose CD-Raft, a protocol designed for distributed systems in cross-domain sites. By introducing two new strategies—\textit{Fast Return} and \textit{Optimal Global Leader Position}—CD-Raft effectively reduces read/write latency while ensuring cross-domain disaster tolerance.
    Additionally, we verified the correctness of CD-Raft in a formal specification using the TLA+ specification.}
    \item{We implemented a CD-Raft-based prototype system and conducted extensive experiments emulating the YCSB benchmark to validate its performance advantages. Empirical results show that CD-Raft significantly reduces the latency of handling client requests in cross-domain sites.}

\end{itemize}




\section{Background And Motivation}
\label{sectio2}
\begin{figure}[t]
  \centering
  \includegraphics[width=0.4\textwidth]{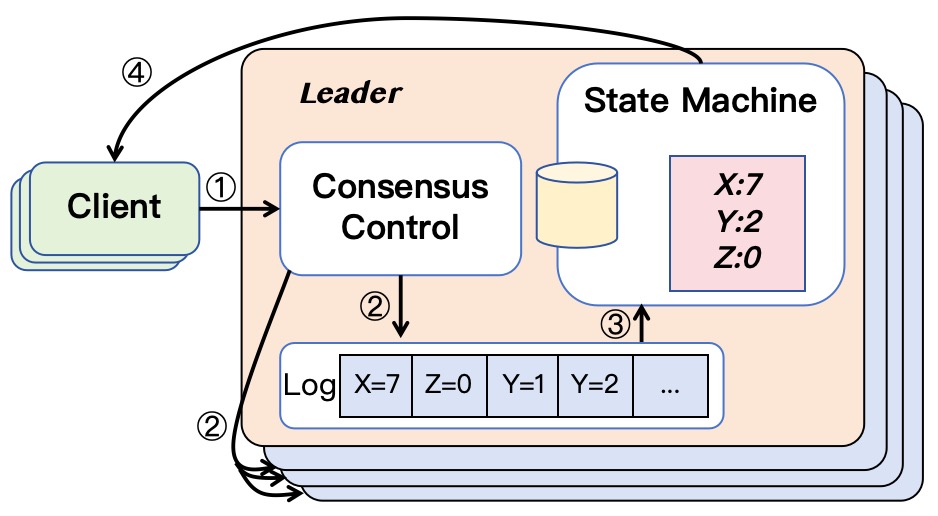} 
   \caption{The write flow of Raft.} 
   \label{writeRaft}
  \vspace{-0.5cm}
\end{figure}

\subsection{Background}\label{background}
Distributed consensus protocols are divided into leaderless consensus protocols and leader-based consensus protocols.
Leaderless consensus protocols, such as EPaxos\cite{EPaxos}, need to determine inter-request dependencies, thus lacking generality.
Furthermore, when requests conflict, they require additional RTTs to resolve those conflicts.
Therefore, we mainly discuss leader-based consensus protocols.
Among them, Multi-Paxos\cite{multipaxos} and Raft\cite{ezpro}, \cite{index} are relatively classic representatives.
Since Raft and Multi-Paxos are similar in design and implementation, we will take Raft as an example below to introduce the processing flow of client read and write requests.

Raft is generally considered easier to understand and deploy than Paxos\cite{paxos1998}, making it a more accessible and implementable alternative.
Since its introduction in 2014, Raft has been rapidly adopted by various real-world distributed systems, including Etcd\cite{etcd}, CockroachDB\cite{cock}, YugabyteDB\cite{yuga}, TiDB\cite{tidb}, Kudu\cite{kudu}, NebulaGraph\cite{NebulaGraph}, LogCabin\cite{Logcabin}, Dragonboat\cite{Dragonboat}, PolarFS\cite{polarFS} and PolarDB\cite{polardb}, etc.
The maintenance of consistency in Raft consists of three core parts: leader election, log replication, and safety assurance.

Raft processes all client requests by electing a unique leader node and ensures linearizability\cite{linear} through three phases: \textit{Append}, \textit{Commit}, and \textit{Apply}.
To facilitate consistent and ordered request processing, Raft employs multiple indexes, including a \textit{commit index}, an \textit{apply index}, and a \textit{read index}.

Fig.~\ref{writeRaft} illustrates the operation flow of a client write request.
When a client sends a write request, it initially reaches the Raft's leader (Step \ding{172}).
The leader appends the request to its local persistent log and forwards it to the followers (Step \ding{173}), waiting for them to complete the log append and acknowledge success.
Upon receiving acknowledgment from a majority of nodes (i.e., more than half of the nodes), the leader marks the request as committed and updates the \textit{commit index}, ensuring data safety even in the event of node failure.
Finally, the leader and followers apply the committed request to their respective state machines (Step \ding{174}) and update their respective \textit{apply index}.
Once the leader completes the apply operation, it returns the result to the client (Step \ding{175}).
This entire process requires that operations be executed in the order of request arrival to maintain data consistency.

\begin{figure}[t]
  \centering
  \includegraphics[width=0.5\textwidth]{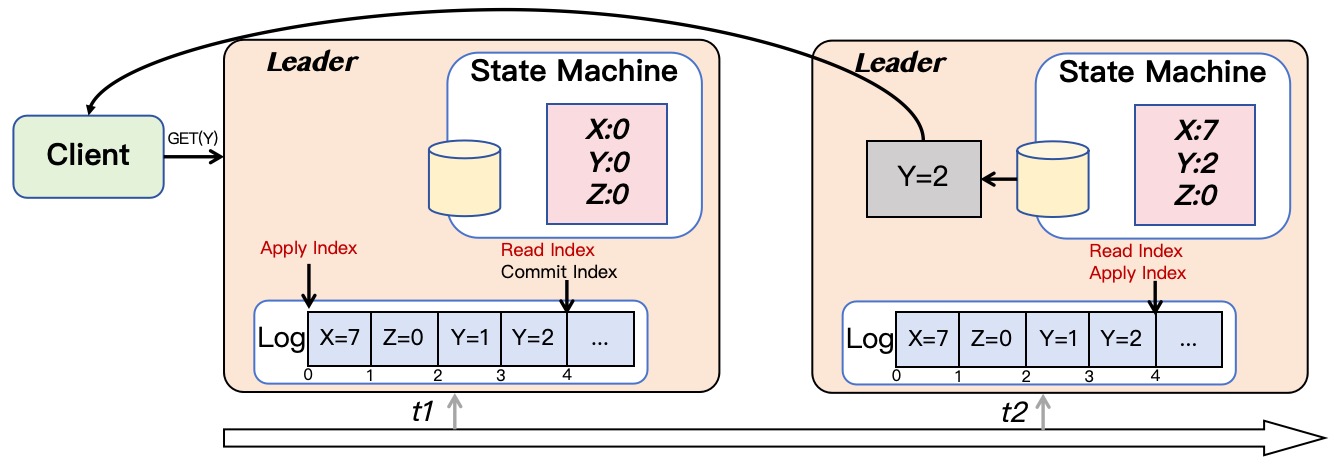} 
   \caption{The read   flow of Raft.} 
   \label{readRaft} 
  \vspace{-0.6cm}
\end{figure}

Fig.~\ref{readRaft} illustrates the operation flow of a client read request\cite{index}.
When the client sends a read request, the leader receives the read request and sets the \textit{read index} to the current \textit{commit index}, i.e., 4.
However, at time \textit{t1}, the leader’s \textit{apply index} remains at 0.
To ensure linearizability, it is necessary to wait until the \textit{apply index} catches up with the \textit{read index} before reading data from the state machine (at time \textit{t2}) to process the Get(Y) operation.
This approach ensures that new data is read instead of out-of-date data, i.e., $''$Y = 2$''$ instead of $''$Y = 0$''$.

\subsection{Motivation}\label{motivation}

In cross-domain distributed systems, maintaining consistency in support of cross-domain disaster tolerance presents a challenge for leader-based consensus protocols due to the high latency involved, particularly when the client and the leader are located in different domains.

\begin{figure}[hbt]
  \centering
  \includegraphics[width=0.5\textwidth]{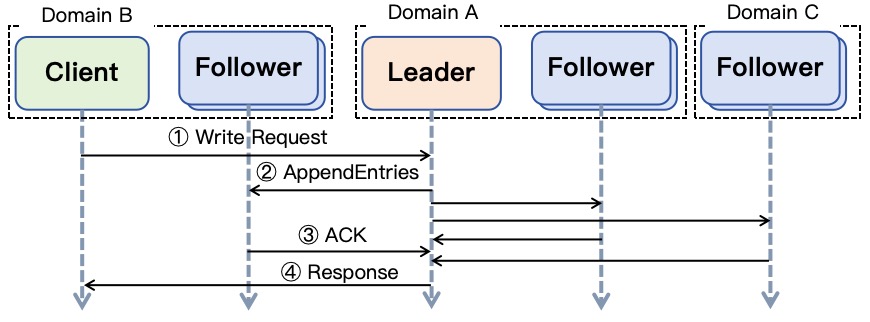} 
   \caption{Write request processing flow of leader-based consensus protocol in cross-domain sites.} 
   \label{newleaderbased}
   
\end{figure}

As illustrated in Fig.~\ref{newleaderbased}, long latency issues arise when the client and the leader node are in separate domains (with the leader in domain A and the client in domain B).
The entire write process requires two cross-domain RTTs (long RTTs).
The first cross-domain RTT occurs between the client and the leader (Step \ding{172} and Step \ding{175}) for the sending and acknowledgment of the request.
The second cross-domain RTT is performed between the leader and nodes in other domains to achieve data consistency confirmation (Step \ding{173} and Step \ding{174}).
\vspace{-0.2cm}
The cross-domain issue significantly increases system response time, particularly under write-intensive workloads, resulting in noticeable waiting times for clients and negatively impacting overall system performance and user experience.
To address this issue, we propose an optimized Raft protocol, termed CD-Raft (Cross-Domain Raft), which builds on the leader-based consensus protocol.
CD-Raft reduces the number of cross-domain RTTs required for write operations to a single RTT through the \textit{Fast Return} strategy, thereby reducing the latency of leader-based consensus protocols in cross-domain distributed systems (see Section \ref{fr} for details).

In addition, in leader-based consensus protocols, the leader is usually elected randomly.
However, due to the differences in the number of client requests in different domains and the uneven communication latency between domains, the position of the leader has a significant impact on the total global latency.
To solve this problem, CD-Raft introduces an \textit{Optimal Global Leader Position} strategy, which can optimize the leader's position to minimize total global latency (see Section \ref{optimal} for details).

\section{Design Of CD-Raft}
\label{sectio3}
To address the above challenges in cross-domain sites, we propose CD-Raft, an optimized Raft protocol specifically for cross-domain sites.
CD-Raft aims to reduce the processing latency of cross-domain requests.

\subsection{Overview}


\vspace{-0.2cm}
\begin{figure}[htp]
    \hspace{-0.2cm} 
    \includegraphics[width=0.5\textwidth]{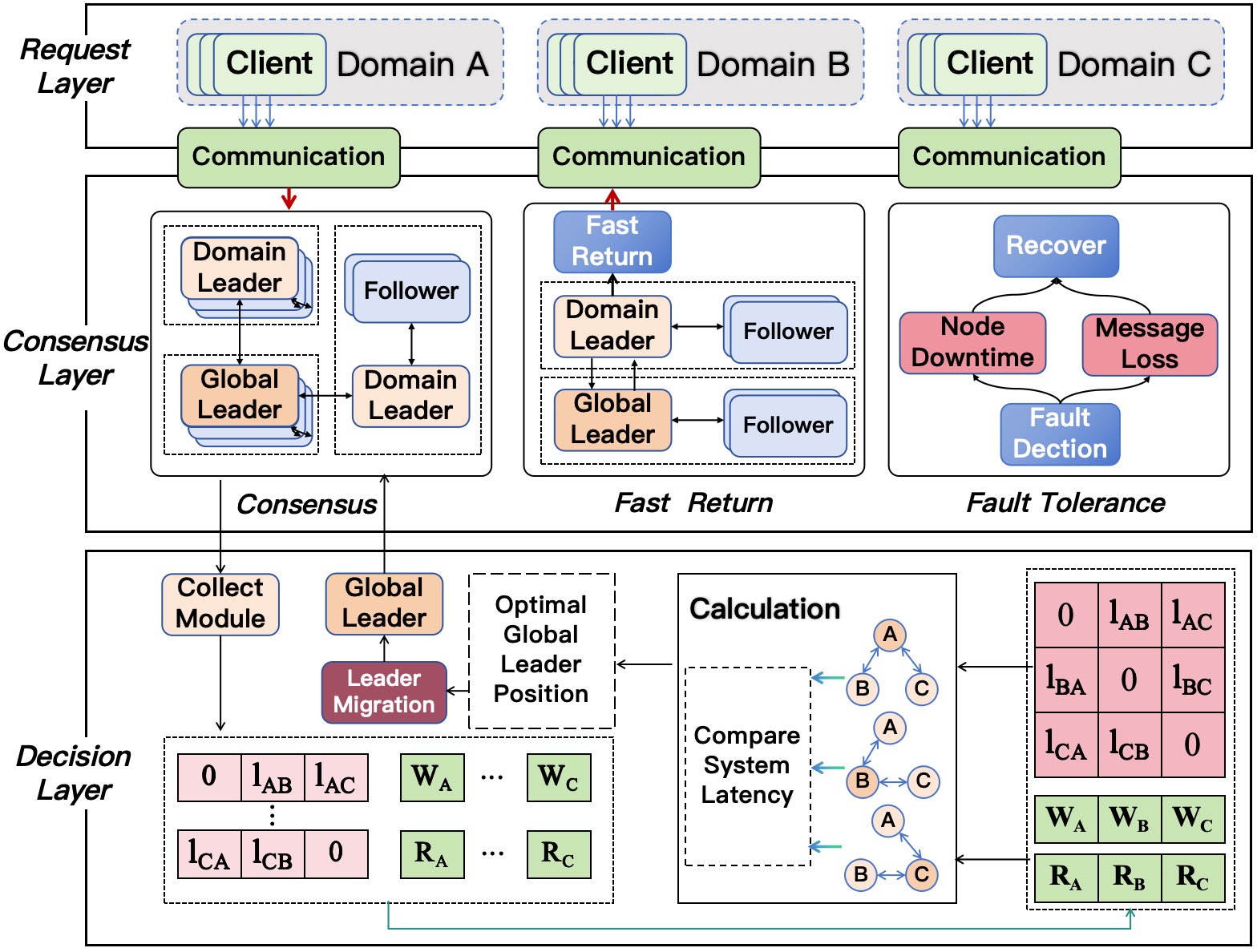}
   \caption{The architecture of CD-Raft.} 
  \label{architecture}
  \centering
  \vspace{-0.3cm}
\end{figure}
\vspace{-0.1cm}
As illustrated in Fig.~\ref{architecture}, CD-Raft adopts a layered architectural design.
It performs the consensus operation upon receiving client requests from the request layer, reducing response times through the \textit{Fast Return} strategy.
CD-Raft offers fault tolerance; when node downtime or message loss occurs, the consensus layer provides appropriate strategies to ensure system availability.
In addition, the decision layer continuously monitors the number of requests from different domains and periodically performs calculations in the background to determine the optimal \textit{Global Leader}, aiming to minimize the total global latency.
CD-Raft effectively reduces system latency while ensuring data security  (i.e., more than half of the nodes in the domain hold data) across at least two domains.
\vspace{-0.1cm}
\subsection{Fast Return}
\label{fr}
As discussed in Section \ref{motivation}, leader-based consensus protocols can experience significant network latency during write operations in cross-domain sites. 
Through the \textit{Fast Return} strategy, this approach reduces the write operation to a single cross-domain RTT.

\begin{figure}[hbt]
  \centering
  
  \includegraphics[width=0.45\textwidth]{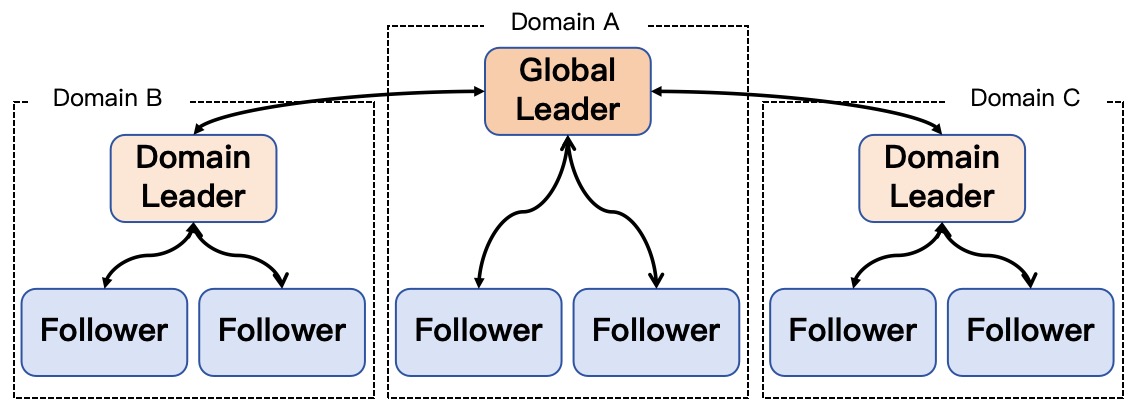} 
  \caption{The communication flow of CD-Raft.} 
  \label{cdraftflow}
  \vspace{-0.4cm}
\end{figure}

Unlike Raft, CD-Raft employs two types of leaders: the \textit{Domain Leader} and the \textit{Global Leader}.
As illustrated in Fig.~\ref{cdraftflow}, CD-Raft elects a \textit{Domain Leader} for each domain and a \textit{Global Leader} from among the \textit{Domain Leaders}. 
The \textit{Global Leader} oversees communication with \textit{Domain Leaders}, while each \textit{Domain Leader} manages consensus operations within its respective domain. 
The \textit{Global Leader} also participates in the consensus process within its domain. 
In-domain consensus operates similarly to the Raft protocol. 
Once a majority of nodes in the domain hold the new data, it is recognized as secure, and the \textit{Domain Leader} communicates this status to the \textit{Global Leader}.

\begin{figure}[htbp]
    \hspace{-0.3cm}
    \centering
    \begin{subfigure}{0.24\textwidth}
        \centering
        \includegraphics[width=\textwidth]{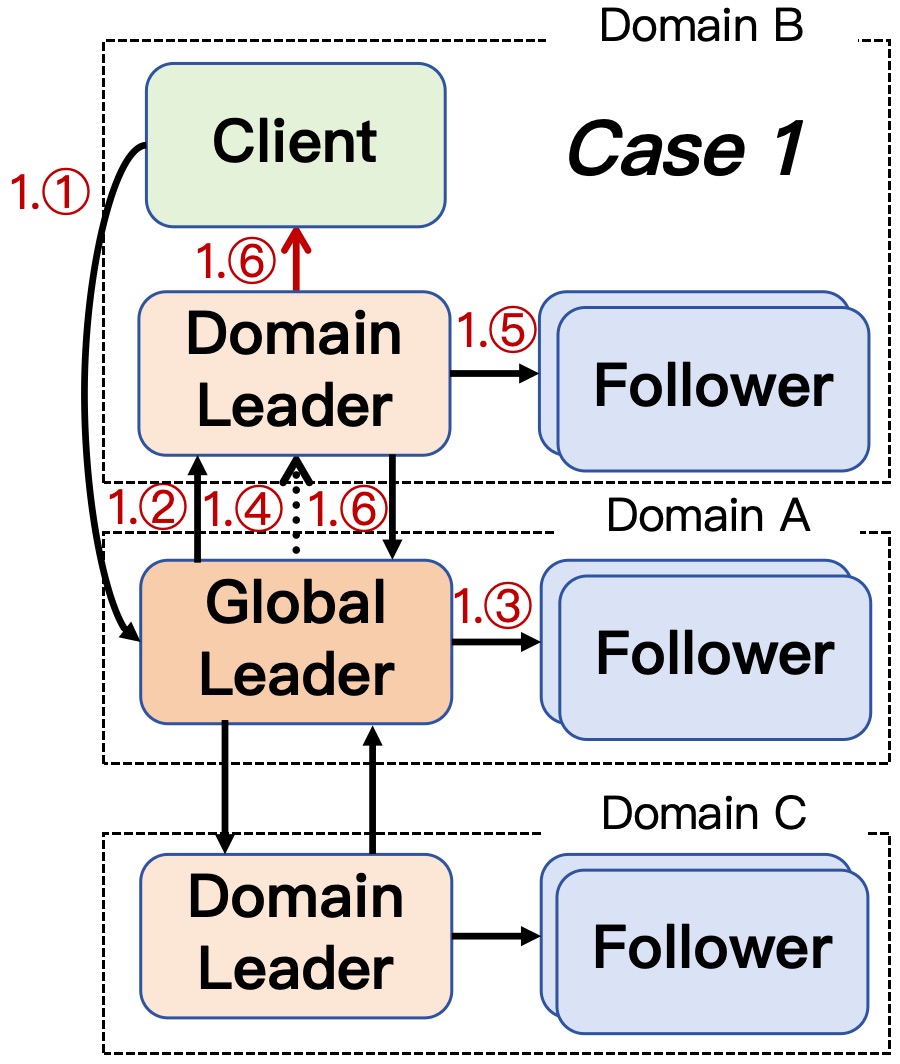}
        \caption{The communication process when the client and the \textit{Global Leader} are in different domains.}
        \label{different}
    \end{subfigure}%
    \hfill
    \begin{subfigure}{0.24\textwidth}
        \centering
        \includegraphics[width=\textwidth]{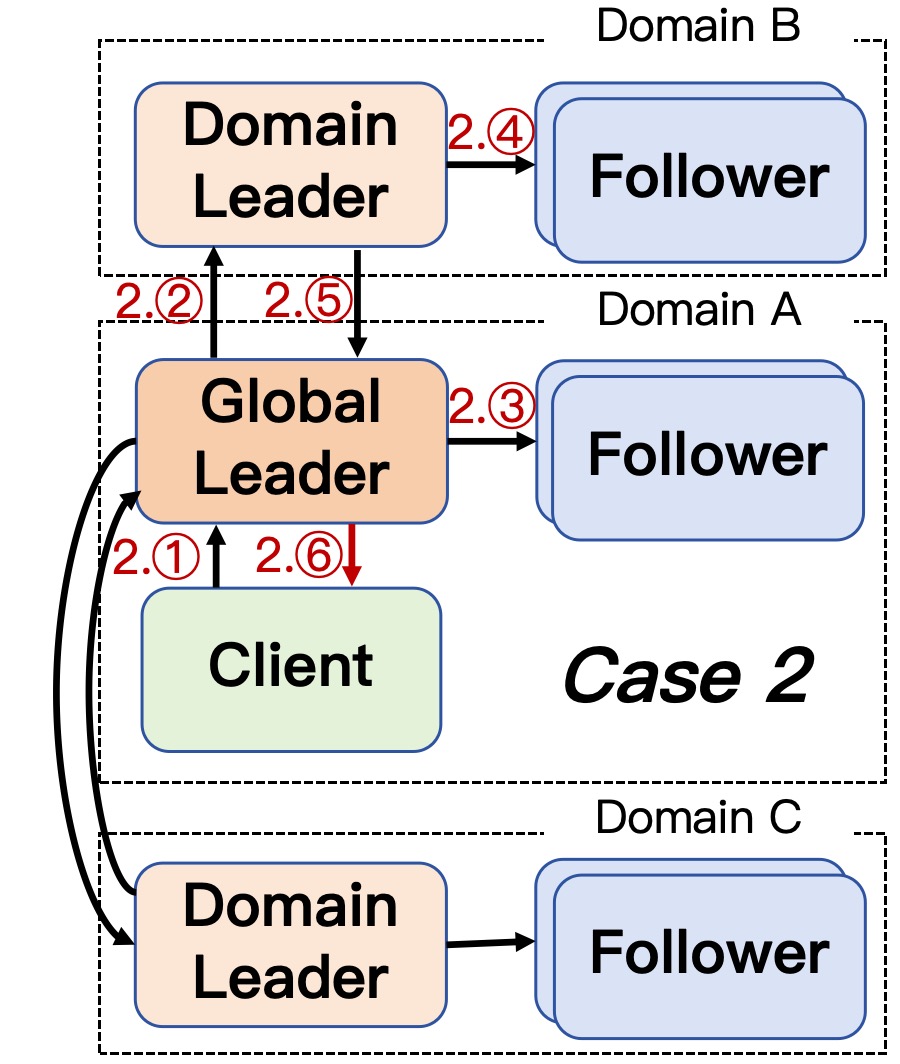}
        \caption{The communication process when the client and the \textit{Global Leader} are in the same domain.}
        \label{same}
    \end{subfigure}%
    \hfill
    \caption{Two communication cases in CD-Raft cross-domain sites.}
    \vspace{-0.3cm}
\end{figure}

When the client and the \textit{Global Leader} are located in different domains, as illustrated in Fig.~\ref{different} (with the client in domain B and the \textit{Global Leader} in domain A), the client’s write request is first sent to the \textit{Global Leader} (Step 1.\ding{172}).
The \textit{Global Leader} then distributes the request to all \textit{Domain Leaders} (Step 1.\ding{173}) and simultaneously manages consensus within its domain (Step 1.\ding{173} and Step 1.\ding{174} in parallel).
Once the majority of nodes in domain A have the latest data, \textit{Global Leader}  sends a confirmation to the \textit{Domain Leader} of domain B (Step 1.\ding{175}).
After receiving the request from the \textit{Global Leader} to replicate the data, the \textit{Domain Leader} will start the consensus in its domain (Step 1.\ding{176}). 
When the majority of nodes in domain B have also synchronized the latest log entries and received acknowledgment from the \textit{Global Leader}, the \textit{Domain Leader} of domain B will directly respond to the client and send an acknowledgment to the \textit{Global Leader} (Step 1.\ding{177}). 
The \textit{Global Leader}, upon receiving this acknowledgment, updates its \textit{commit index} (when the \textit{Global Leader} domain completes the consensus operation and receives consensus acknowledgment from another domain).
Since the client and the \textit{Domain Leader} of domain B are within the same domain, this acknowledgment process is swift and not counted within the cross-domain RTT.
Steps 1.\ding{173} and 1.\ding{176} occur simultaneously with Steps 1.\ding{174} and 1.\ding{175}, involving the same cross-domain network latency and log appending Steps.
This design ensures that the cross-domain network transmission latency of the write operation approximates a single cross-domain RTT (equivalent to the duration of Step 1.\ding{172} and Step 1.\ding{173} combined) while simultaneously guaranteeing data security across at least two domains (domains A and B).


When the client and the \textit{Global Leader} are located in the same domain (both in domain A), as shown in Fig.~\ref{same}, the client’s request is sent to the \textit{Global Leader} (Step 2.\ding{172}).
Upon receiving the request, the \textit{Global Leader} distributes the request to all \textit{Domain Leaders} (Step 2.\ding{173}) while concurrently managing log appending within its domain (Step 2.\ding{174}). 
After receiving the request, each \textit{Domain Leader} starts the consensus in its domain (Step 2.\ding{175}).
After receiving acknowledgment from the other domain (Step 2.\ding{176}) and ensuring that the majority of the nodes in its domain hold the new data, the \textit{Global Leader} responds to the client (Step 2.\ding{177}).
This approach guarantees data security across at least two domains.
In this case, the cross-domain RTT is typically equivalent to the RTT between the \textit{Global Leader}’s domain and the nearest other domain.


  

The read operation of CD-Raft is similar to Raft (see Section \ref{background}), when the client's read request reaches the \textit{Global Leader}, the \textit{Global Leader} sets the value of the \textit{read index} of the read request to the value of the \textit{Global Leader}'s \textit{in-domain commit index} (only the \textit{Global Leader} domain reaches consensus log index), and when the \textit{apply index} of the \textit{Global Leader} catches up with the \textit{read index}, it can execute the read operation, read the corresponding data from the state machine, and respond to the client.

\subsection{Safety}
CD-Raft's commit condition is that data can be committed once it is held by the majority of nodes in both the \textit{Global Leader}’s domain and another domain.
Consequently, the \textit{Global Leader} election requirements are more stringent.
To ensure safety, a non-empty intersection must exist between the set of nodes participating in leader election and the set of nodes involved in subsequent consensus decisions\cite{flexible}.

Assuming there are N domains, CD-Raft’s \textit{Global Leader} election requires at least N-1 \textit{Domain Leaders} to participate in the election, and the new \textit{Global Leader}’s log must be more up-to-date than those of N-2 other \textit{Domain Leaders}.
This stringent requirement ensures that brain-split does not occur, even in the event of network partition.
For \textit{Domain Leader} election, because data is still required to be held by a majority of nodes within the domain, the election still follows the majority principle.
\vspace{-0.2cm}
\subsection{Fault Tolerance}
Since node failures and network errors are inevitable, distributed systems inevitably face various problems. 
To ensure the normal operation of the system, it is essential to have fault tolerance. 
System failures are mainly divided into two categories: node downtime issues and message loss issues. 
The following will provide CD-Raft fault tolerance solutions for these two types of failures.

\subsubsection{\textbf{Node Fault Tolerance}}
As node downtime is inevitable, certain failures occur within the system. 
The node failures that CD-Raft can tolerate are as follows.
\begin{itemize}
    \item If no more than half of the nodes in the domain containing the Global Leader are down, and at least one other domain retains over half of its nodes, CD-Raft can continue to provide service to clients. 
    \item Even if more than half of the nodes in the Global Leader's domain are down, CD-Raft can still function as long as no more than two domains have more than half of their nodes down simultaneously.
\end{itemize}

\begin{figure}[!t]
    \vspace{-0.3cm}
    \centering
    \begin{subfigure}{0.23\textwidth}
        \centering
        \includegraphics[width=\textwidth]{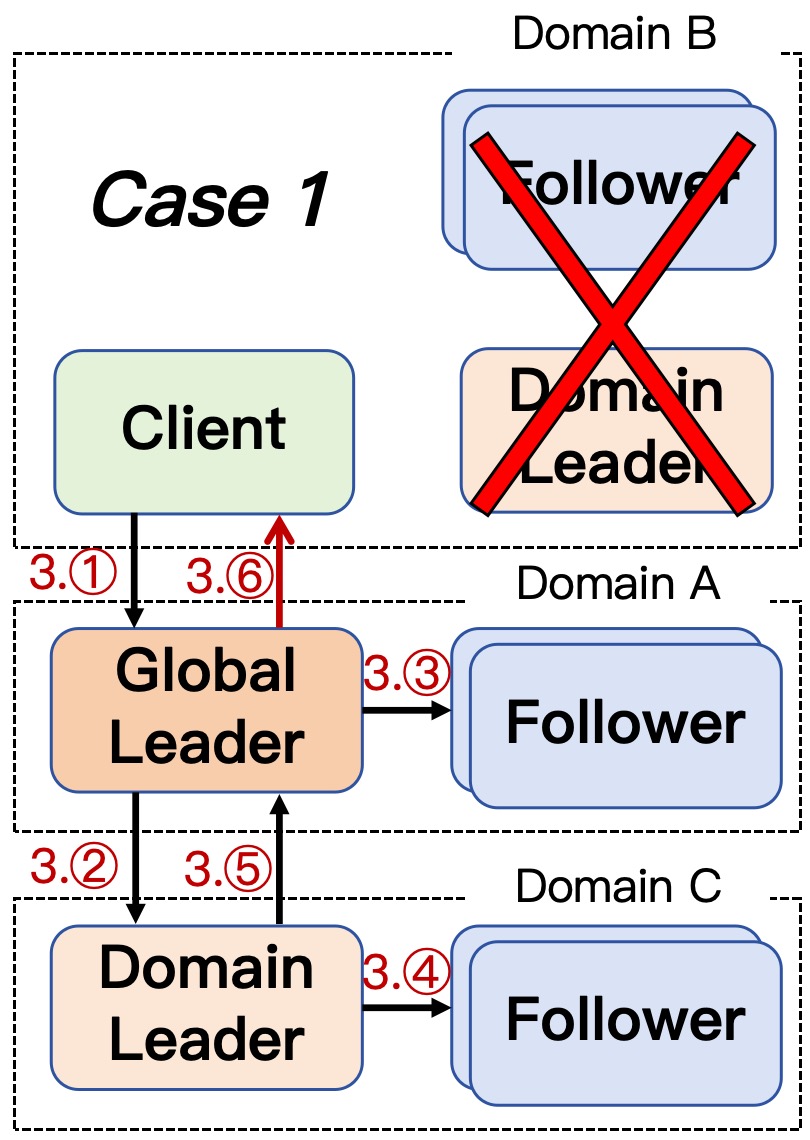}
        \caption{Failure scenarios for non-Global Leader domains.}
        \label{f1}
    \end{subfigure}%
    \hfill
    \begin{subfigure}{0.23\textwidth}
        \centering
        \includegraphics[width=\textwidth]{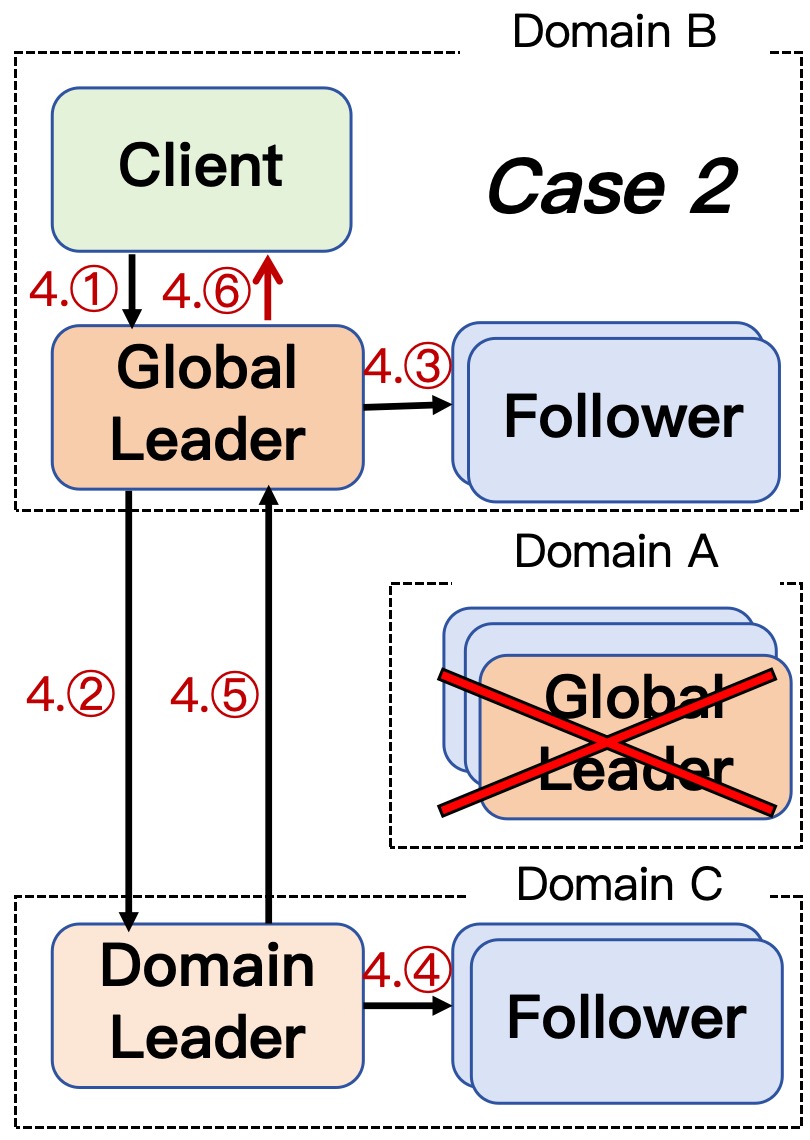}
        \caption{Failure scenarios for Global Leader domain.}
        \label{f2}
    \end{subfigure}%
    \hfill
    \caption{CD-Raft fault tolerance in cross-domain failure sites.}
    \vspace{-0.6cm}
\end{figure}

In the first failure scenario, consider a three-domain system, as shown in Fig.~\ref{f1}, where the \textit{Global Leader} is located in domain A. 
If domain B becomes unavailable (with more than half of its nodes down) and the client is located in domain B, the \textit{Global Leader} takes over the role of \textit{Domain Leader} in domain B to send an acknowledgment to the client. 
Additionally, the \textit{Global Leader} ensures that a majority of nodes in at least one other domain besides it hold the latest data. 
This setup still ensures data security across at least two domains. 
When the client is in domain A or domain C, the client request processing flow remains unchanged.

In the second failure scenario, when more than half of the nodes in the \textit{Global Leader}'s domain fail, this results in the unavailability of that domain, necessitating a new round of Global Leader election. 
As shown in Fig.~\ref{f2}, if domain A becomes unavailable and the log of the \textit{Domain Leader} in domain B is newer than domain C, the \textit{Domain Leader} of domain B takes on the role of the new \textit{Global Leader}. 
If the client is in domain B, CD-Raft processes the read/write request as described in the scenario where the client and \textit{Global Leader} reside in the same domain, requiring a single cross-domain RTT (Step 4.\ding{173} and Step 4.\ding{176}). 
If the client is in domain C, CD-Raft still employs the \textit{Fast Return} strategy, with the \textit{Domain Leader} in domain C sending the acknowledgment to the client. 
If the client is in domain A, this failure case is managed as described in the first failure scenario above.

\subsubsection{\textbf{Message Fault Tolerance}}
Due to the complexity of network environments, message loss is inevitable. Using Fig.~\ref{different} as an example, we illustrate CD-Raft's handling strategy in cases of network message loss.

If \textbf{Message \ding{172}} is lost, the client does not receive a response within the expected time, resulting in a request timeout and automatic retransmission.

If \textbf{Message \ding{173}}, where the \textit{Global Leader} communicates with the \textit{Domain Leader} in domain B, is lost, but the \textit{Global Leader} can still communicate with another domain (e.g., domain C), CD-Raft handles the situation according to the strategy illustrated in Fig.~\ref{f1}. When the \textit{Global Leader} cannot communicate with any other domain, the client experiences a timeout and retransmits the request due to the lack of acknowledgment.

If \textbf{Message \ding{174}} between the \textit{Global Leader} and more than half of the followers in the domain is lost, the client does not receive a response, and the request is retransmitted after a timeout.

If \textbf{Message \ding{175}} is lost, CD-Raft is unable to execute the \textit{Fast Return} strategy. When the \textit{Global Leader} receives an acknowledgment from another domain, it performs the commit and responds to the client.

If \textbf{Message \ding{176}} between the \textit{Domain Leader} and more than half of the followers in the domain is lost, domain B cannot reach a consensus, preventing the \textit{Domain Leader} from executing the \textit{Fast Return} strategy, similar to the scenario of node downtime. If other \textit{Domain Leaders} complete consensus on the new request and respond to the \textit{Global Leader}, the \textit{Global Leader} performs the commit and responds to the client.

If \textbf{Message \ding{177}} sent by the \textit{Domain Leader} of domain B to the \textit{Global Leader} is lost, the \textit{Domain Leader} can still respond directly to the client, provided that at least two domains have the latest data. However, since the \textit{Global Leader} has not received acknowledgment from domain B, its \textit{commit index} cannot be updated. The \textit{Global Leader} can only perform the commit after receiving acknowledgment from another domain. If a read request arrives, since the \textit{read index} is set to the \textit{Global Leader}'s \textit{in-domain commit index}, the \textit{Global Leader} waits for acknowledgment from the other domain before executing the commit. Once the commit is completed and the state machine is applied, the read request is executed, and the client receives the data.
\vspace{-0.1cm}
\subsection{Optimal Global Leader Position}
\label{optimal}



Due to varying client request volumes and inter-domain latencies, the position of the \textit{Global Leader} impacts total global latency.
In order to obtain the \textit{Global Leader} position that minimizes the total global latency, we need to evaluate total global latency based on the client request volume distribution and the latency between domains. Table~\ref{model} below summarizes the symbols used in the following sections.
\vspace{-0.2cm}
\begin{table}[h]
    \centering
    \caption{Symbols used in the system performance model.} 
    \label{model}
    \begin{tabular}{lp{0.74\columnwidth}} 
        \toprule
        Symbols & Descriptions \\ 
        \midrule
        $l_{ij}$ & Latency between Domain $i$ and Domain $j$ \\ 
        $L_{i}$  & Cross-domain latency for read/write operations by clients in Domain $i$ \\ 
        $W_{i}$  & Number of write requests by clients in Domain $i$ \\ 
        $R_{i}$  & Number of read requests by clients in Domain $i$ \\ 
        $L_{s}$  & Total global latency in the distributed system \\ 
        $N$      & Number of domains \\ 
        \bottomrule
    \end{tabular}
    \vspace{-0.2cm}
\end{table}

When the read/write request volume from different domains (i.e., $W_{i}$ and $R_{i}$) and the inter-domain latency (i.e., $l_{ij}$) are known, we only need to find a position for the \textit{Global Leader} that minimizes the total cross-domain latency, thereby maximizing system performance. Assume the cluster is distributed across $N$ domains, with the \textit{Global Leader} node located in domain $z$ ($z \in [1, N]$). The cross-domain latency $L_{i}$ for client requests from domain $i$ ($i \in [1, N]$) falls into three cases.

When domain $i$ is available and the client is in the same domain as the \textit{Global Leader} (i.e., $i = z$), the \textit{Global Leader}, upon receiving a write request, forwards the request to all \textit{Domain Leaders} and initiates its own in-domain consensus process. As soon as the \textit{Global Leader} receives an acknowledgment from any other domain, it can respond to the client. This cross-domain latency is typically the latency between the \textit{Global Leader}’s domain and the nearest other domain. Since read requests do not involve cross-domain latency, $L_{i}$ satisfies the conditions of Equation (1). Let $j$ represent another available domain.
\begin{equation}
    L_i = 2 \cdot \min(l_{zj}) \cdot W_i 
    \tag{1}
\end{equation}

When domain $i$ is available and the client is in a different domain from the \textit{Global Leader} (i.e., $i \neq z$), the \textit{Global Leader}, upon receiving a write request, forwards the request to all \textit{Domain Leaders} and initiates its in-domain consensus process. Once the \textit{Domain Leader} in the domain where the client resides successfully commits, it can directly respond to the client. Therefore, the write request experiences a cross-domain RTT. The client’s read request is sent directly to the \textit{Global Leader}, and after reading the data from the \textit{Global Leader}, it is immediately returned to the client. This entire read process also experiences a cross-domain RTT, which is the same as the cross-domain RTT for the write process. Thus, $L_{i}$ is as shown in Equation (2).
\begin{equation}
    L_i = 2 \cdot l_{iz} \cdot (W_i + R_i) \tag{2}
\end{equation}

When domain $i$ is unavailable, since the Global Leader must reside in an available domain, it follows that $i \neq z$. The \textit{Global Leader}, upon receiving a write request, forwards the request to all \textit{Domain Leaders} and initiates its in-domain consensus process. Once the \textit{Global Leader} receives an acknowledgment from any other domain, it can respond to the client. This cross-domain latency is typically the latency between the \textit{Global Leader}'s domain and the nearest other available, communicable domain. The client’s read request is sent directly to the \textit{Global Leader}, and after the data is read from the \textit{Global Leader}, it is immediately returned to the client. This entire read process also experiences a cross-domain RTT. Thus, $L_{i}$ is as shown in Equation (3), where domain $j$ is an available domain.
\begin{equation}
    L_i = 2 \cdot (l_{iz} + \min(l_{zj})) \cdot W_i + 2 \cdot l_{iz} \cdot R_i  \tag{3}
\end{equation}

Thus, the total global latency is as shown in Equation (4). We only need to find a Global Leader domain (i.e., domain $z$) that minimizes $L_{s}$ to reduce the system’s global cross-domain latency to the greatest extent possible.
\begin{equation}
    L_s = \sum_{i=1}^{N} L_i \qquad i \in [1, N]
    \tag{4}
\end{equation}

The cost of \textit{Global Leader} migration is a cross-domain RTT, so the time of periodic judgment ($Tp$) needs to be much greater than the cross-domain RTT. 
When the \textit{Global Leader} is in the $z$ domain, Equation (5) should be satisfied.
\begin{equation}
    Tp \gg 2 \cdot \max(l_{zj})
    \tag{5}
\end{equation}

\subsection{Linearizability of CD-Raft}
CD-Raft ensures linearizability because its operations for appending, committing, and applying logs are strictly executed in index order, and CD-Raft satisfies the following two properties \cite{Harmonia}, \cite{bucraft}, sufficient to guarantee linearizability.

\textit{\textbf{P1.Visibility.}} A read operation sees the effects of all write operations that finished before it started.

\textit{\textbf{P2.Integrity.}} A read operation does not see the effects of any write operation that had not been committed at the time the read finished.

In CD-Raft, when the \textit{Global Leader} processes a read operation, it sets the \textit{read index} to the \textit{in-domain commit index} of the \textit{Global Leader}. Because the \textit{commit index} is less than or equal to the \textit{in-domain commit index}, the read operation can see all the write operations that have been committed, thus satisfying P1. In addition, because read requests can only see the results of write operations after \textit{Apply}, and \textit{Apply} can only be done after \textit{Commit}, so read operations cannot see write operations that have not been committed, thus satisfying P2. 

\subsection{Correctness}
We developed a formal specification and correctness proof for the CD-Raft described in this section. 
The formal specification, written in the TLA+ specification language\cite{specifying}, precisely models the \textit{Fast Return} strategy under a dual-leader architecture.  
We have mechanically proved the correctness of CD-Raft using the TLA+ proof system \cite{TLA+}.
The TLA+ source code is available at: https://github.com/Czq0001/CD-Raft-TLA.git.

\section{Implementation And Evaluation}
\label{sectio4}

\subsection{System Implementation}
\label{implementation}
We have implemented a CD-Raft prototype system in Go language, employing RocksDB\cite{rockdb} as the underlying key-value storage engine.
The distributed communication framework employs gRPC\cite{grpc} and Google Protocol Buffers\cite{protobuf}, enabling efficient node-to-node data serialization and message exchange. 
\begin{table}[t] %
    \centering
    \caption{The YCSB workloads used in evaluations.} 
    \label{ycsb}    
    \begin{tabular}{lllc}
        \toprule
        \textbf{Workload} & \textbf{Write Type} & \textbf{Query Type} & \textbf{Category} \\
        \midrule
        Load & Insert & / & Insert Only \\
        A    & Update & Point Query & 50\%update 50\%read \\
        B    & Update & Point Query & 5\%update 95\%read \\
        C    & /      & Point Query & Read Only \\
        \bottomrule
    \end{tabular}
    \vspace{-0.5cm}
\end{table}

\begin{figure*}[htbp]
    \centering
    \begin{subfigure}{0.33\textwidth}
        \centering
        \includegraphics[width=\textwidth]{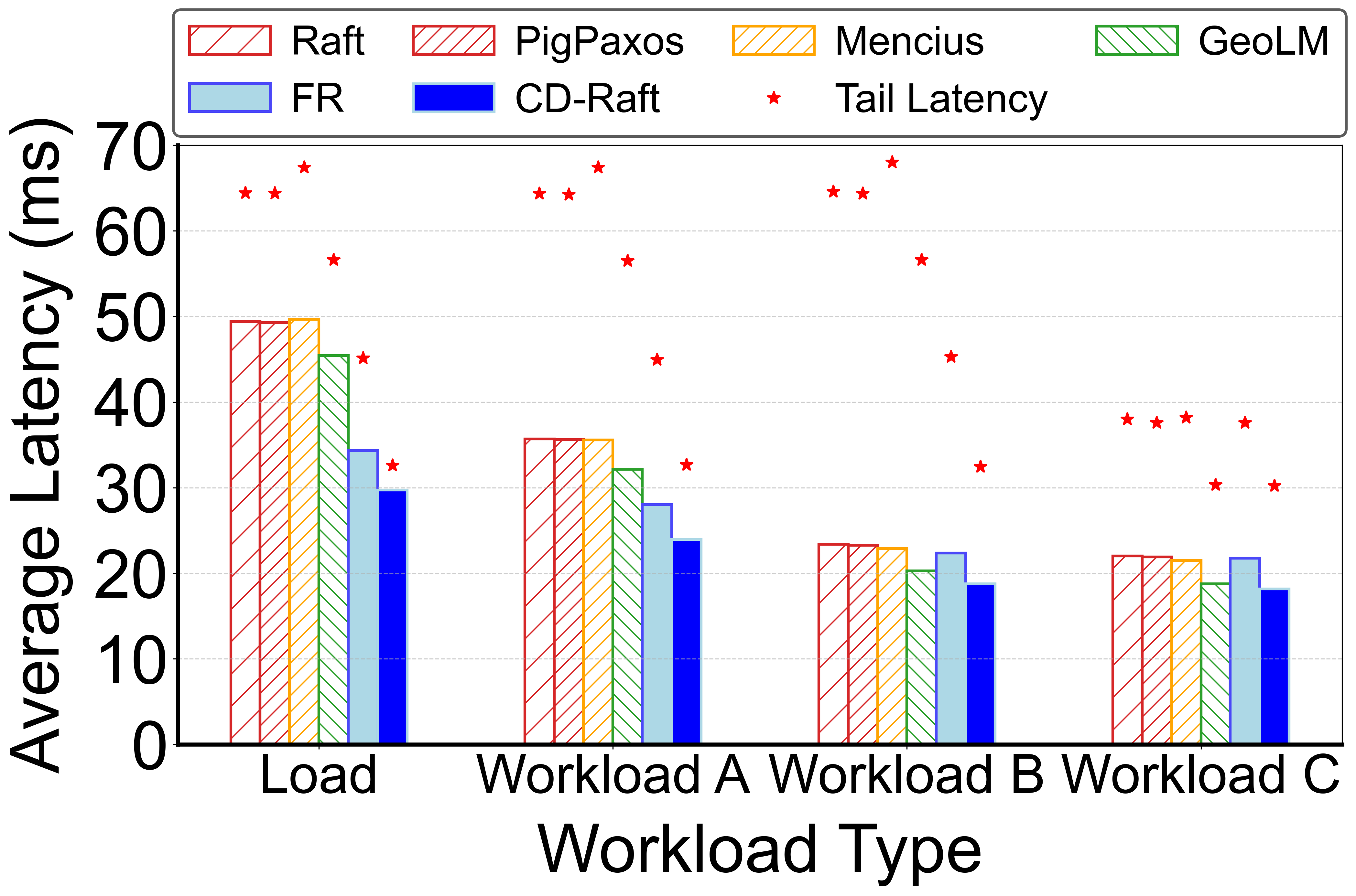}
        
        \caption{Average latency}
        \label{overalla}
    \end{subfigure}%
    \hfill 
    \begin{subfigure}{0.33\textwidth}
        \centering
        \includegraphics[width=\textwidth]{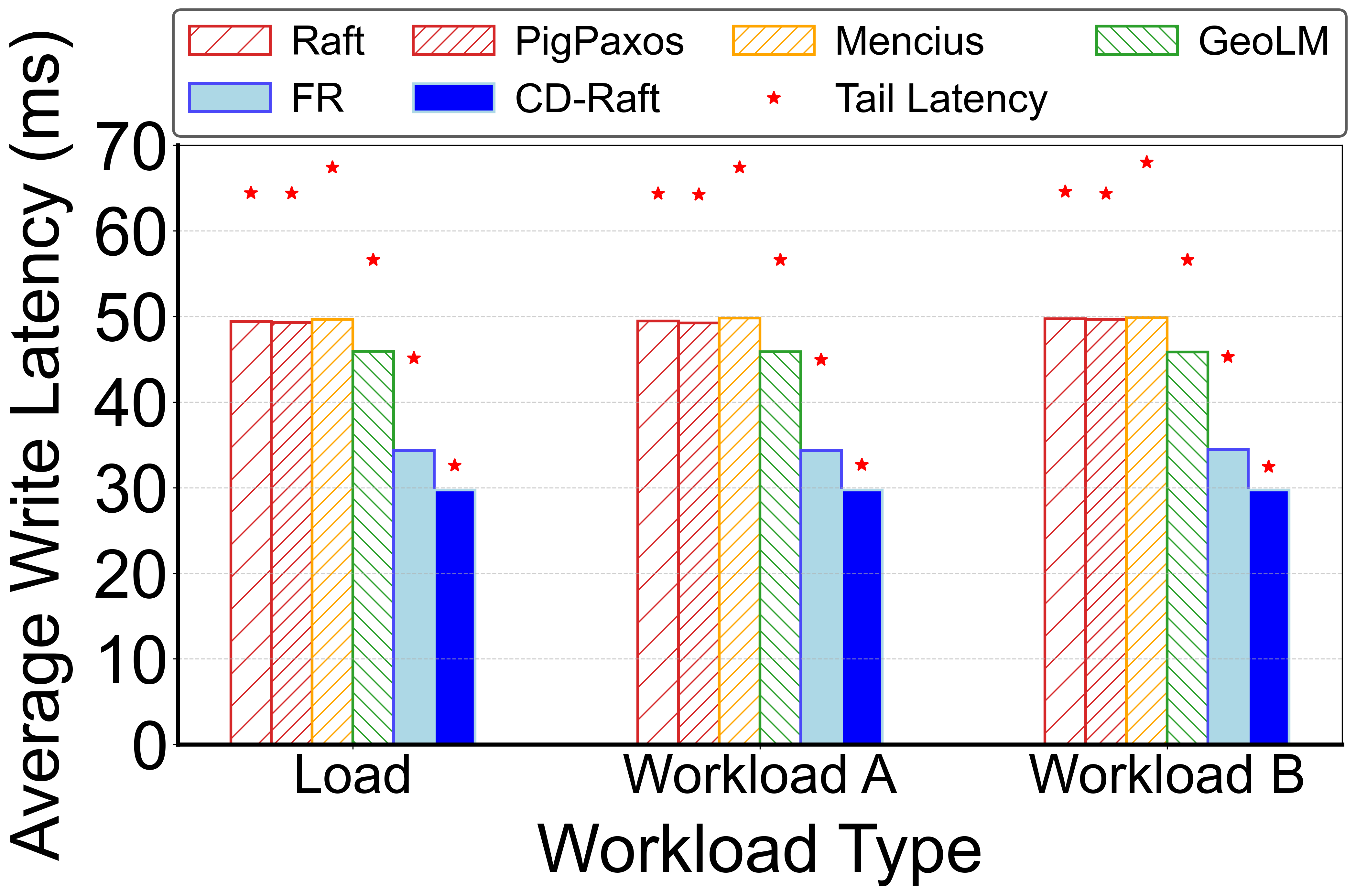}
        
        \caption{Average write latency}
        \label{overallb}
    \end{subfigure}%
    \hfill
    \begin{subfigure}{0.33\textwidth}
        \centering
        \includegraphics[width=\textwidth]{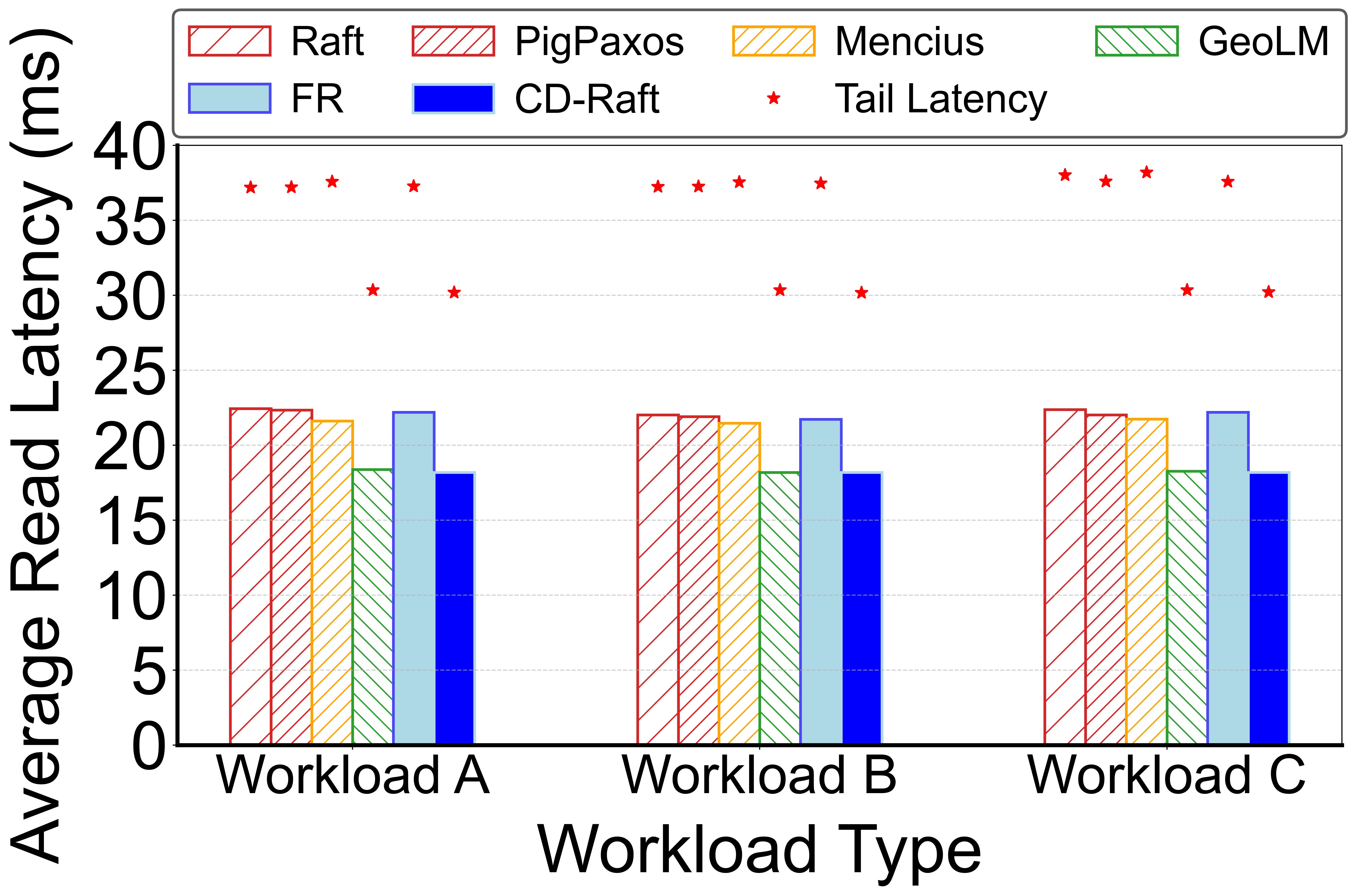}
        
        \caption{Average read latency}
        \label{overallc}
    \end{subfigure}
    \caption{Comparison of latencies among Raft, PigPaxos, Mencius, GeoLM, FR, and CD-Raft under YCSB workloads.}
    \vspace{-0.6cm}
\end{figure*}
\vspace{-0.2cm}
\subsection{Experimental Setup}
\label{setup}
\noindent\textbf{\textit{Environment and Parameters.}} To evaluate the actual performance of CD-Raft, we conducted experiments on the Huawei Cloud Platform.
These experiments were performed in a real cross-domain network environment containing four domains (Beijing, Shanghai, Guangzhou, and Guiyang), with three server nodes and one client node deployed in each domain.
Each node was equipped with 32GB of DRAM and a 1TB SSD, running the Ubuntu 22.04.4 LTS operating system.
Server nodes ran instances of the consensus protocol under evaluation. 
Client nodes utilized YCSB to generate request streams for assessing the performance of CD-Raft.

\textbf{\textbf{WorkLoad.}} To systematically evaluate the performance benefits of CD-Raft, we conducted extensive comparative experiments across various dimensions. 
Detailed workload parameters are shown in Table~\ref{ycsb}. 
These included Load (insert-only), Workload A (50:50 read/update ratio), Workload B (95:5 read/update ratio), and Workload C (read-only). 
Keys followed a Zipf distribution. 
This approach simulates real-world scenarios where certain keys experience higher access frequency. 
Each key-value pair contained a 16 B key and a 1 KB value by default.
$Tp$ is set to 100 times of cross-domain RTT.

\textbf{\textbf{Baselines.}} 
To comprehensively evaluate the performance of CD-Raft, we selected the following six protocols for comparison:
\begin{itemize}
    \item  \textit{Raft:} A widely adopted, leader-based consensus protocol that ensures strong consistency.
    \item \textit{PigPaxos}\cite{pigpaxos}:  A consensus protocol designed to optimize cross-domain communication through a hierarchical architecture.
    \item \textit{Mencius}\cite{mencius}:  A consensus protocol that employs a rotating leader strategy.
    \item \textit{GeoLM}\cite{geolm}: A leader management strategy designed to improve Raft's performance in cross-domain environments by adapting to network and application changes.
    \item \textit{FR}: A basic version of CD-Raft implementing only the \textit{Fast Return} strategy. This version is used to isolate and evaluate the performance gain specifically from this strategy.
    \item \textit{CD-Raft}: The complete CD-Raft system, featuring tightly coupled \textit{Fast Return} and \textit{Optimal Global Leader Position} strategies.
\end{itemize}

\textbf{\textit{Leader Configuration.}}
In this experimental evaluation, we configured the leader locations for the compared protocols as follows:
For the Raft, PigPaxos, and FR, in scenarios that included the Beijing domain, their leader (specifically, the \textit{Global Leader} for FR) was fixed in the Beijing domain; otherwise, it was fixed in the Shanghai domain (see Section~\ref{geographical topology} for details). 
CD-Raft, in contrast, utilized its \textit{Optimal Global Leader Position} strategy to select the optimal \textit{Global Leader} domain. 
Through such a comparative setup, the purpose is to quantify the effects of the \textit{Fast Return} strategy and the \textit{Optimal Global Leader Position} strategy on reducing latency.
The standard three-domain configuration mentioned later refers to the Beijing-Shanghai-Guangzhou domains.

\vspace{-0.1cm}
\subsection{Overall Results}
\label{overall}

\begin{figure*}[tp]
    \centering
    \begin{subfigure}{0.33\textwidth}
        \centering
        \includegraphics[width=\textwidth]{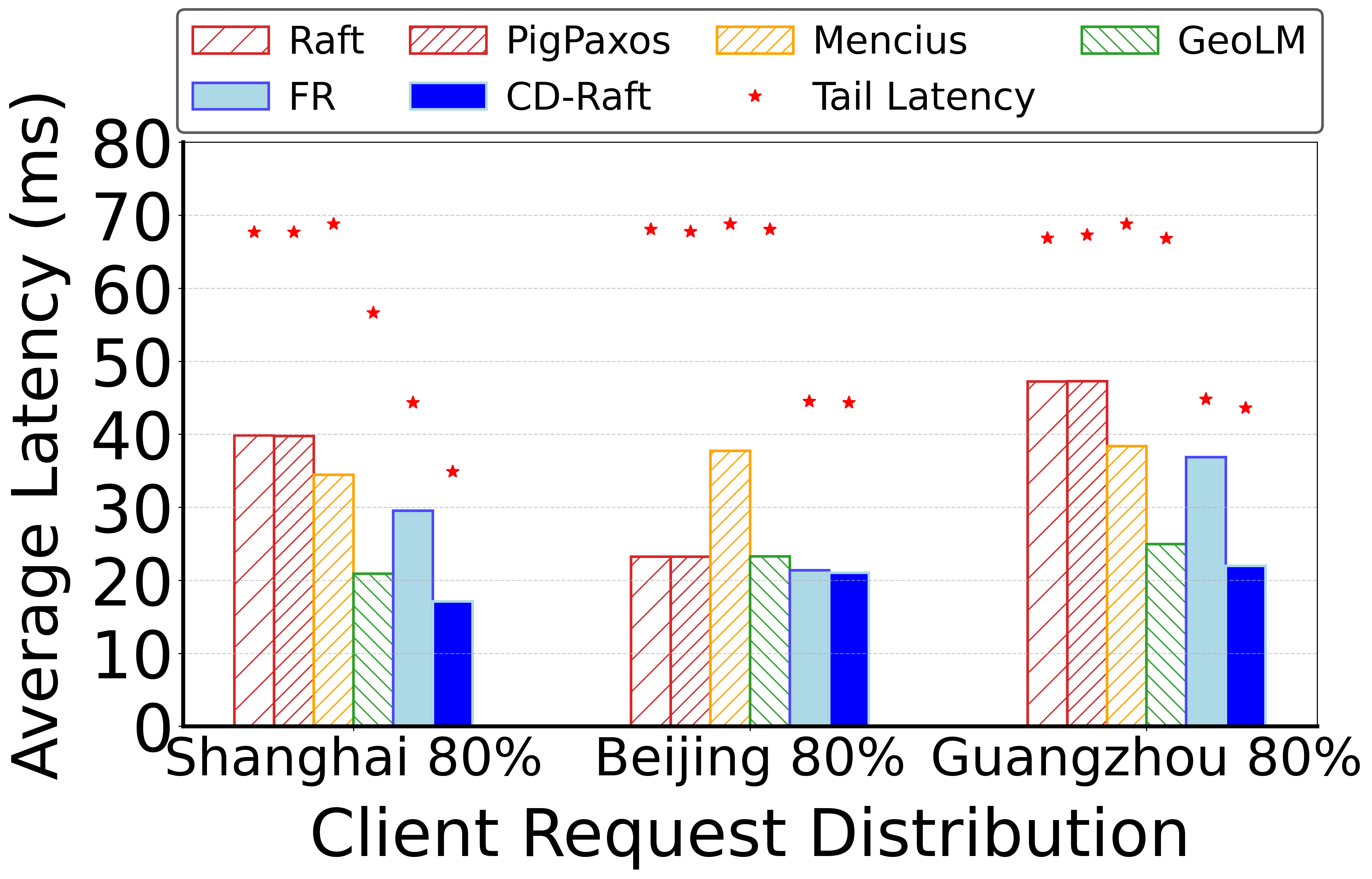}
        \caption{Average latency}
        \label{A8}
    \end{subfigure}%
    \hfill
    \begin{subfigure}{0.33\textwidth}
        \centering
        \includegraphics[width=\textwidth]{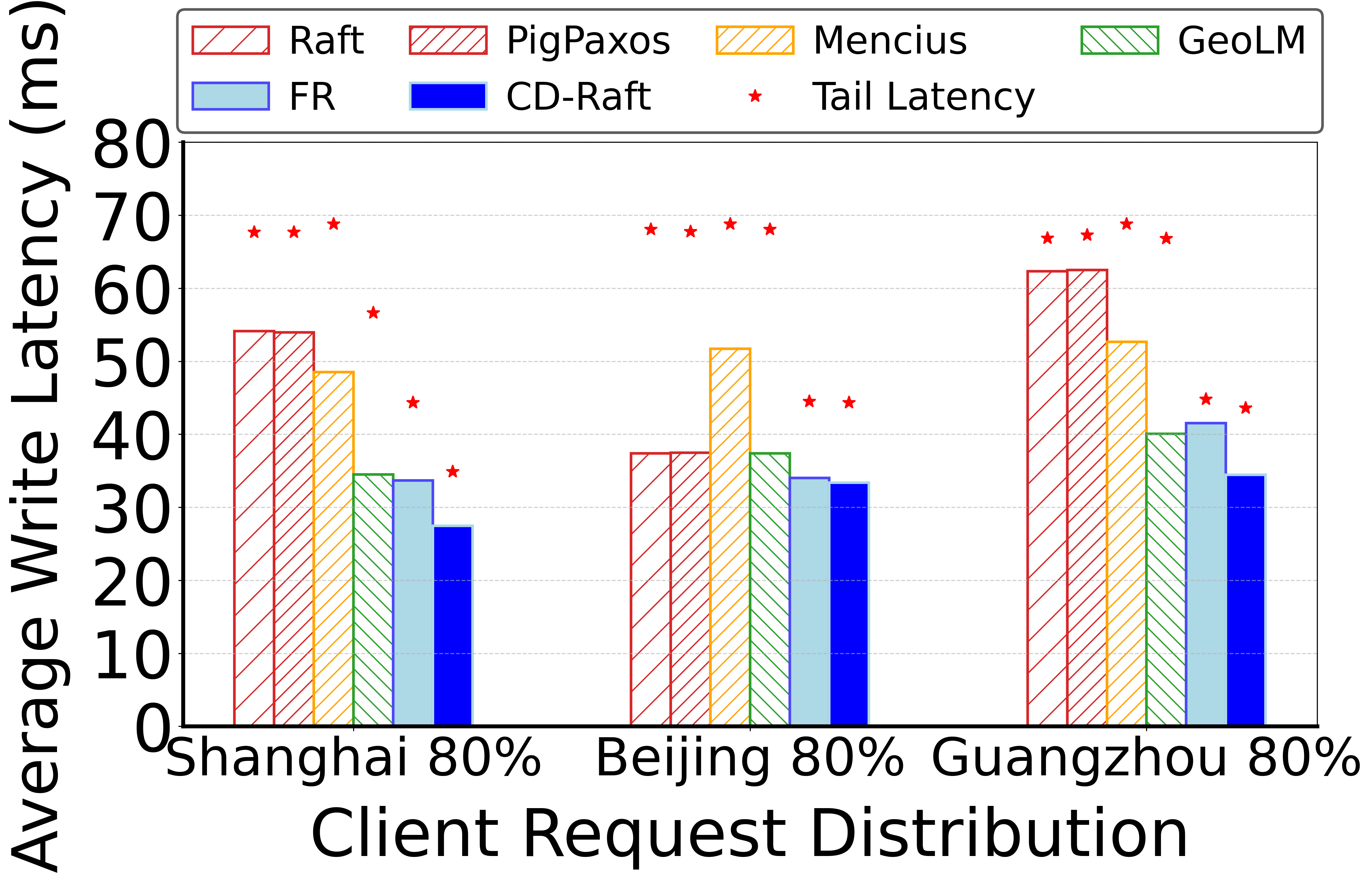}
        \caption{Average write latency}
        \label{W8}
    \end{subfigure}%
    \hfill
    \begin{subfigure}{0.33\textwidth}
        \centering
        \includegraphics[width=\textwidth]{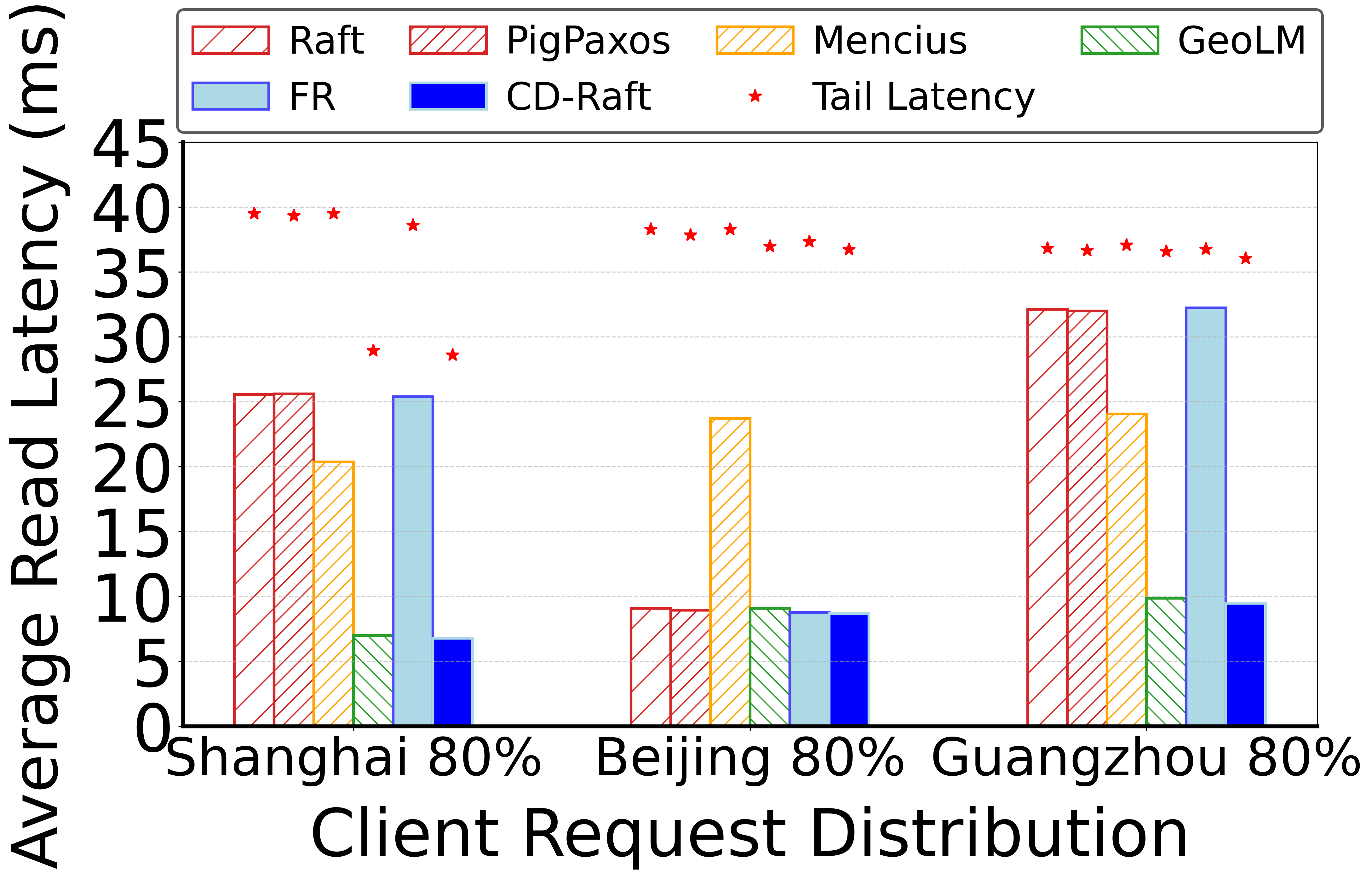}
        \caption{Average read latency}
        \label{R8}
    \end{subfigure}
    \caption{Comparison of latencies among Raft, PigPaxos, Mencius, GeoLM, FR, and CD-Raft under Different Client Request Distributions(8-1-1).}
    \vspace{-0.5cm}
    \label{l8}
\end{figure*}

\begin{figure*}[htbp]
    \centering
    \begin{subfigure}{0.33\textwidth}
        \centering
        \includegraphics[width=\textwidth]{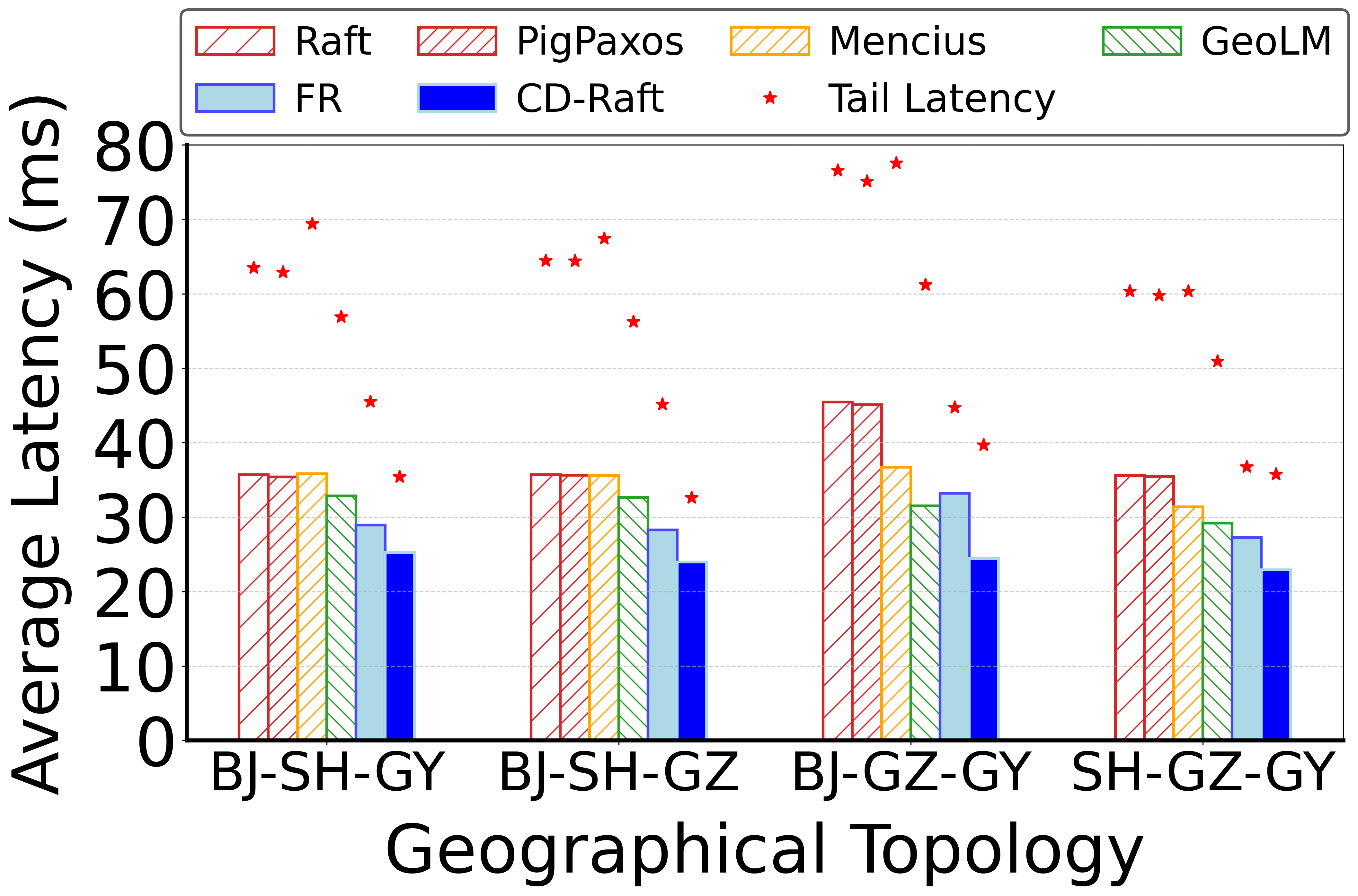}
        \caption{Average latency}
        \label{inerA}
    \end{subfigure}%
    \hfill
    \begin{subfigure}{0.33\textwidth}
        \centering
        \includegraphics[width=\textwidth]{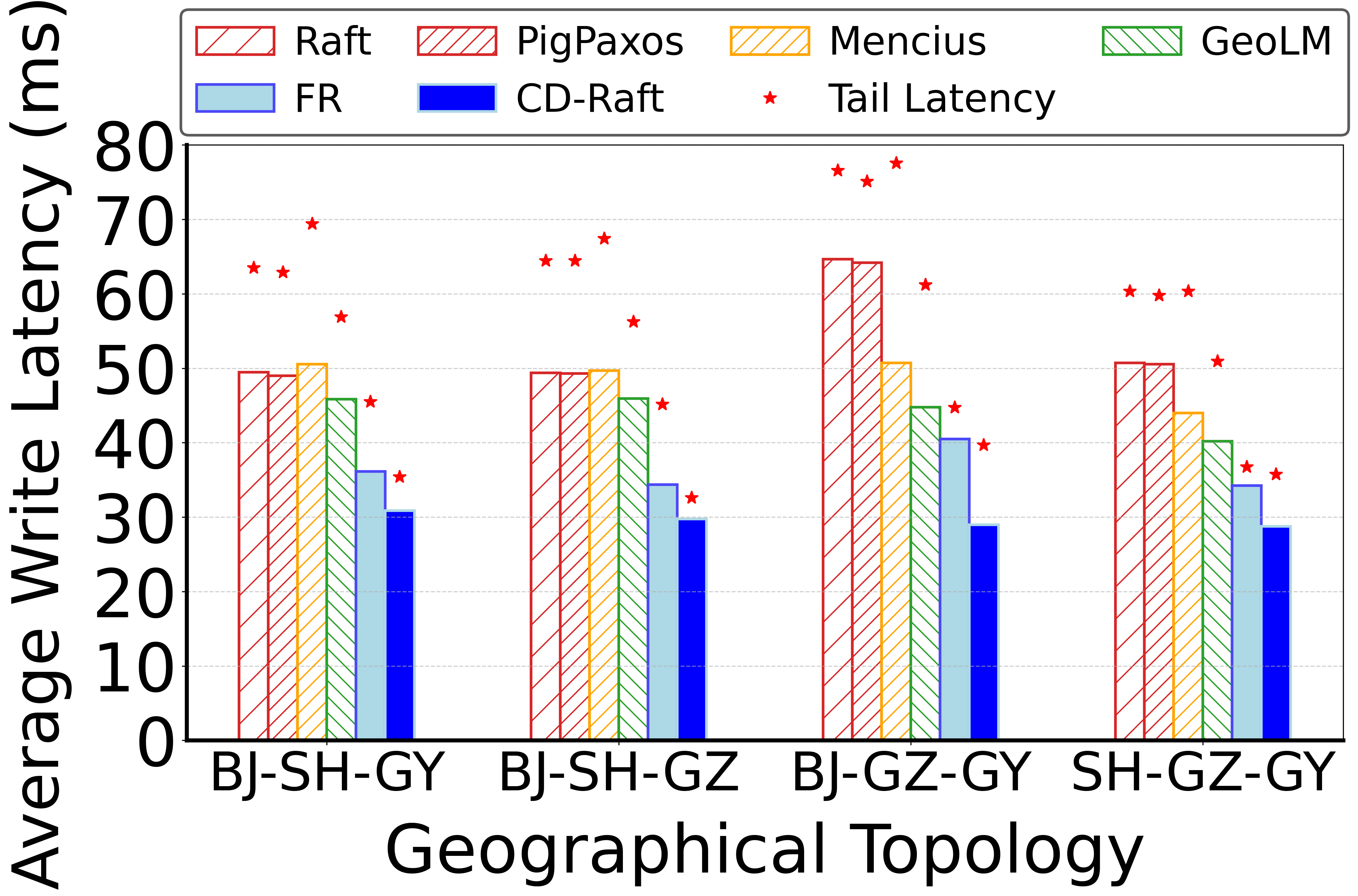}
        \caption{Average write latency}
        \label{inerW}
    \end{subfigure}%
    \hfill
    \begin{subfigure}{0.33\textwidth}
        \centering
        \includegraphics[width=\textwidth]{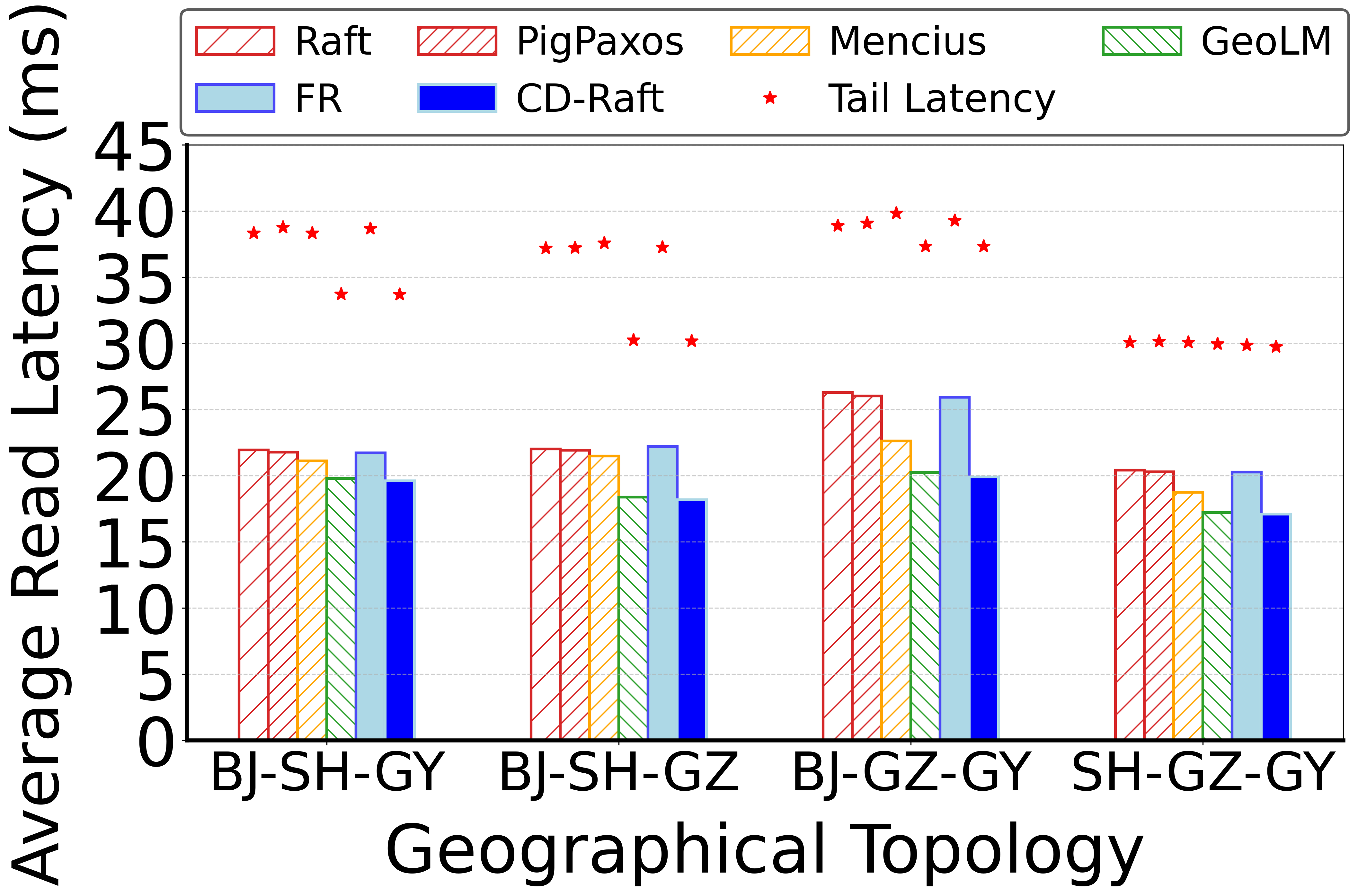}
        \caption{Average read latency}
        \label{inerR}
    \end{subfigure}
    \caption{Comparison of latencies among Raft, PigPaxos, Mencius, GeoLM, FR, and CD-Raft under Different Geographical Topologies.}
    \vspace{-0.5cm}
    \label{num}
\end{figure*}

To evaluate the overall performance of CD-Raft in cross-domain sites, we conducted experiments using a standard three-domain configuration. 
We measured and compared the average operation latency, as well as (99th percentile) tail latency, of CD-Raft against baseline protocols Raft, PigPaxos, Mencius, GeoLM, and FR under four YCSB workloads.
At the beginning of the experiment, each client first performed a data load of 1 million key-value pairs, and then performed 1 million operations for Workload A, Workload B, and Workload C respectively.
\vspace{-1px}

\textbf{\textit{Latency.}} Fig.~\ref{overalla} shows the average latency of each protocol under different YCSB workloads. 
Raft and PigPaxos have very similar performances because their leaders are both fixed in Beijing and their core processes are similar. 
Mencius, as a rotating leader protocol, can be regarded as the average performance of Raft under different domains.
GeoLM is the leader's performance on the optimal domain, but the write request of the client outside the leader domain still needs two cross-domain RTTs.

Under heavy write loads (Load and WorkLoad A), CD-Raft shows significant performance advantages. 
Compared to Raft, PigPaxos, Mencius, and GeoLM, CD-Raft's average latency under the Load workload decreased by 34\%$\sim$41\%, and its tail latency also decreased by 42\%$\sim$52\%.
FR, which only uses the \textit{Fast Return} strategy, also performs well, with its average latency reduced by 24\%$\sim$31\%, and its tail latency also reduced by 20\%$\sim$34\%.
Under WorkLoad A, CD-Raft's average latency decreased by 25\%$\sim$33\%, and its tail latency also decreased by 42\%$\sim$52\% compared with Raft, PigPaxos, Mencius, and GeoLM.

Under heavy read loads (WorkLoad B and WorkLoad C), the improvement in average latency for CD-Raft is relatively limited.
However, even in these read-dominant scenarios, CD-Raft can still significantly reduce tail latency when write requests are present.
Under Workload B, CD-Raft's average latency was reduced by 7\%$\sim$20\% and its tail latency was also reduced by 49\%$\sim$53\%, compared to Raft, PigPaxos, Mencius, and GeoLM.
When there is no write request, GeoLM can achieve similar performance as CD-Raft.




As Fig.\ref{overallb} shows, CD-Raft achieves the lowest average write latency under different YCSB workloads.
The average write latency of CD-Raft is 39\%$\sim$41\% lower than that of Raft, PigPaxos and Mencius, and 35\%$\sim$36\% lower than that of GeoLM. 
Regarding tail latency, CD-Raft reduced tail latency by 49\%$\sim$53\% compared to Raft, PigPaxos, and Mencius, and by 42\%$\sim$43\% compared to GeoLM.
Even when FR's \textit{Global Leader} is not in the optimal domain, it still achieves significant performance improvements.
The average write latency of FR is 30\%$\sim$32\% lower than that of Raft, PigPaxos, and Mencius, and 24\%$\sim$26\% lower than that of GeoLM. 

As Fig.\ref{overallc} shows, CD-Raft achieves the lowest average read latency under different YCSB workloads.
The average read latency of CD-Raft is 17\%$\sim$19\% lower than that of Raft and PigPaxos, and 15\%$\sim$17\% lower than that of Mencius. 
Regarding tail latency, CD-Raft reduced it by 18\%$\sim$21\% compared to Raft, PigPaxos and Mencius.  

The experimental results above clearly show that CD-Raft exhibits significant optimization effects in terms of average latency, write latency, read latency, and tail latency.

\vspace{-0.1cm}
\subsection{Impacts of the Distribution of Client Requests}
\label{load distribution}
To investigate the impact of the geographical distribution of client requests on protocol performance, experiments in this section were based on the standard three-domain configuration. 
We systematically evaluated system performance under a skewed workload distribution across the three domains with an 80\%-10\%-10\% ratio (denoted 8-1-1). 
Each client performed 1 million requests, evenly split between reads and writes (50\% each). 
This evaluation aimed to reveal the sensitivity and adaptability of each protocol to variations in geographical load distribution, with a particular focus on the optimization effectiveness of CD-Raft in these scenarios.

As Fig.\ref{l8} shows, CD-Raft achieves the lowest latency in all cases.
Even when 80\% of requests are in the Beijing domain (where the leader of Raft is located), CD-Raft still reduces average latency by 9.42\% and tail latency by 48.44\% compared to Raft, with both read and write latencies showing respective decreases.



\begin{figure*}[tp]
    \centering
    \begin{subfigure}{0.33\textwidth}
        \centering
        \includegraphics[width=\textwidth]{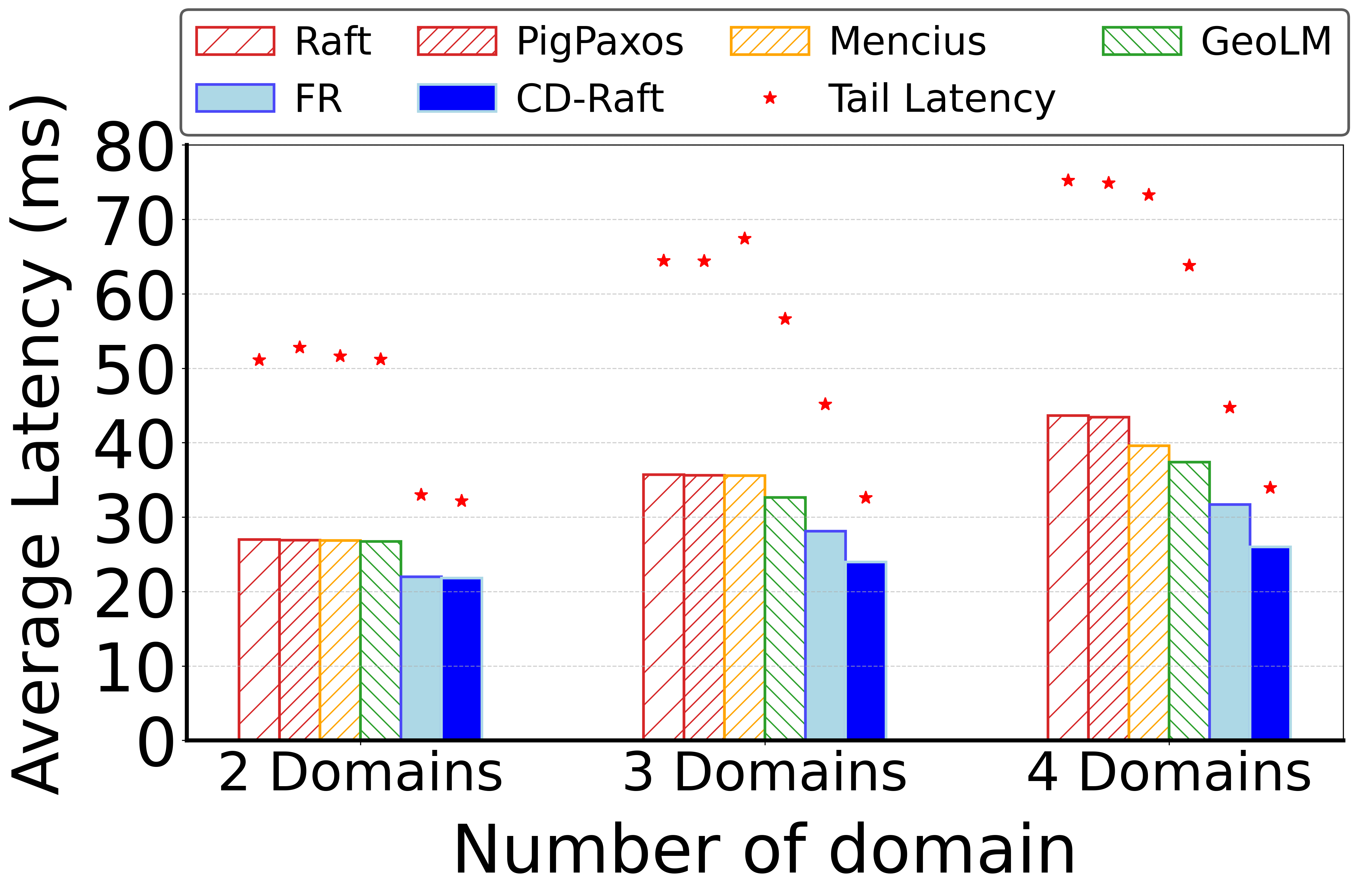}
        \caption{Average latency}
        \label{numA}
    \end{subfigure}%
    \hfill
    \begin{subfigure}{0.33\textwidth}
        \centering
        \includegraphics[width=\textwidth]{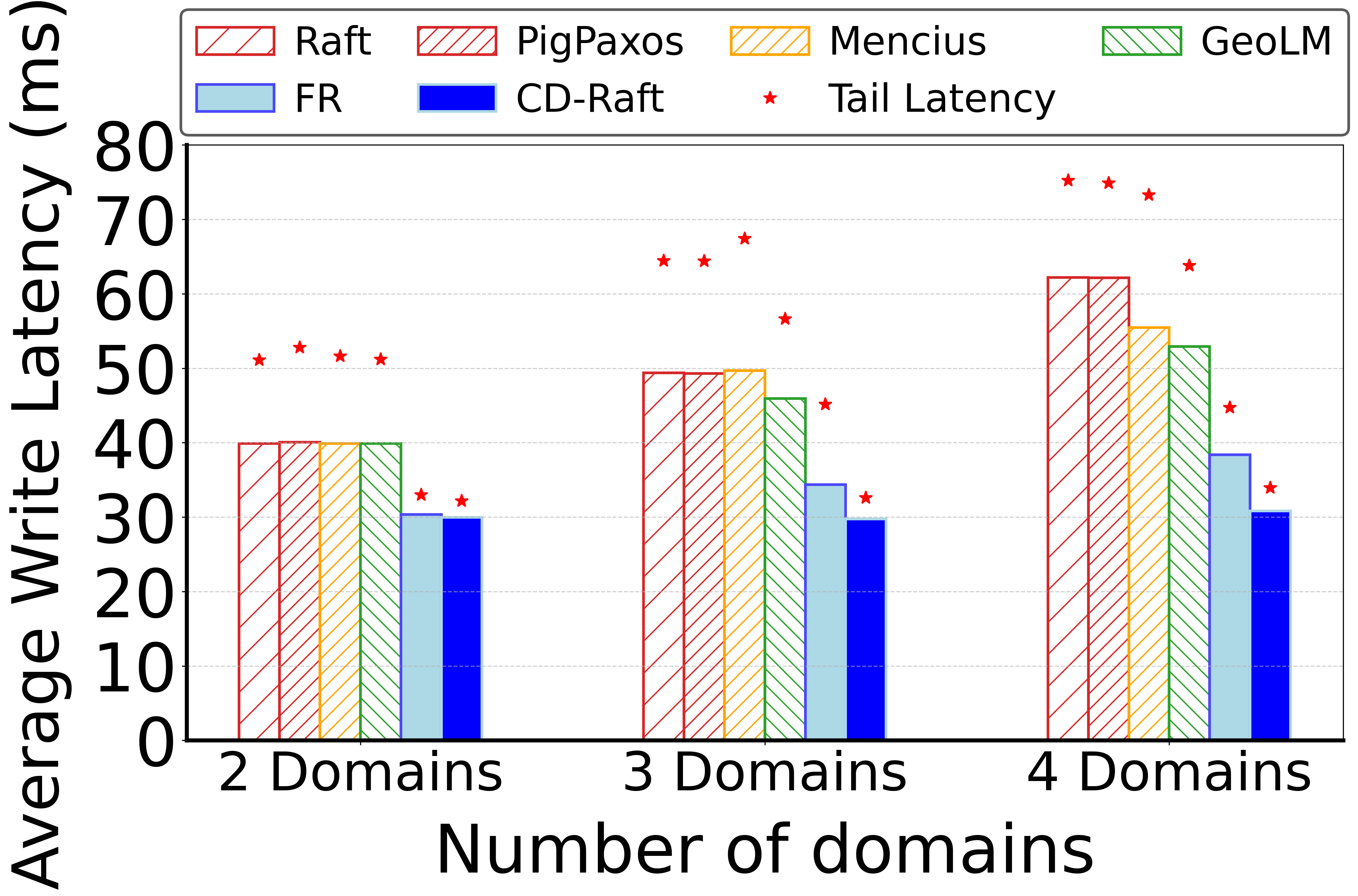}
        \caption{Average write latency}
        \label{numW}
    \end{subfigure}%
    \hfill
    \begin{subfigure}{0.33\textwidth}
        \centering
        \includegraphics[width=\textwidth]{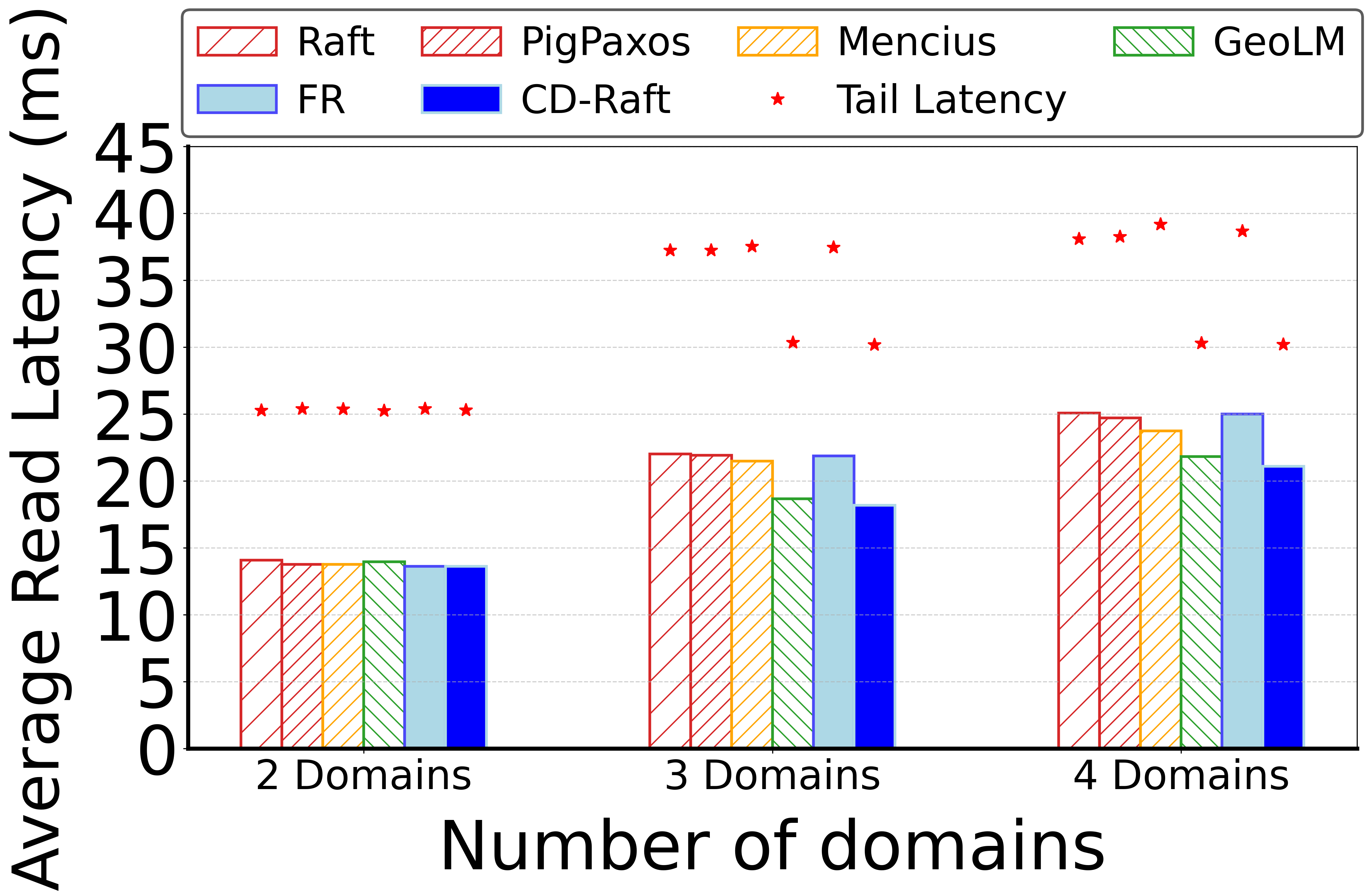}
        \caption{Average read latency}
        \label{numR}
    \end{subfigure}
    \caption{Comparison of latencies among Raft, PigPaxos, Mencius, GeoLM, FR, and CD-Raft with Varying Domain Counts.}
    \label{iner}
    \vspace{-0.6cm}
\end{figure*}

\begin{figure}[t]
    \centering 
    \includegraphics[width=0.48\textwidth]{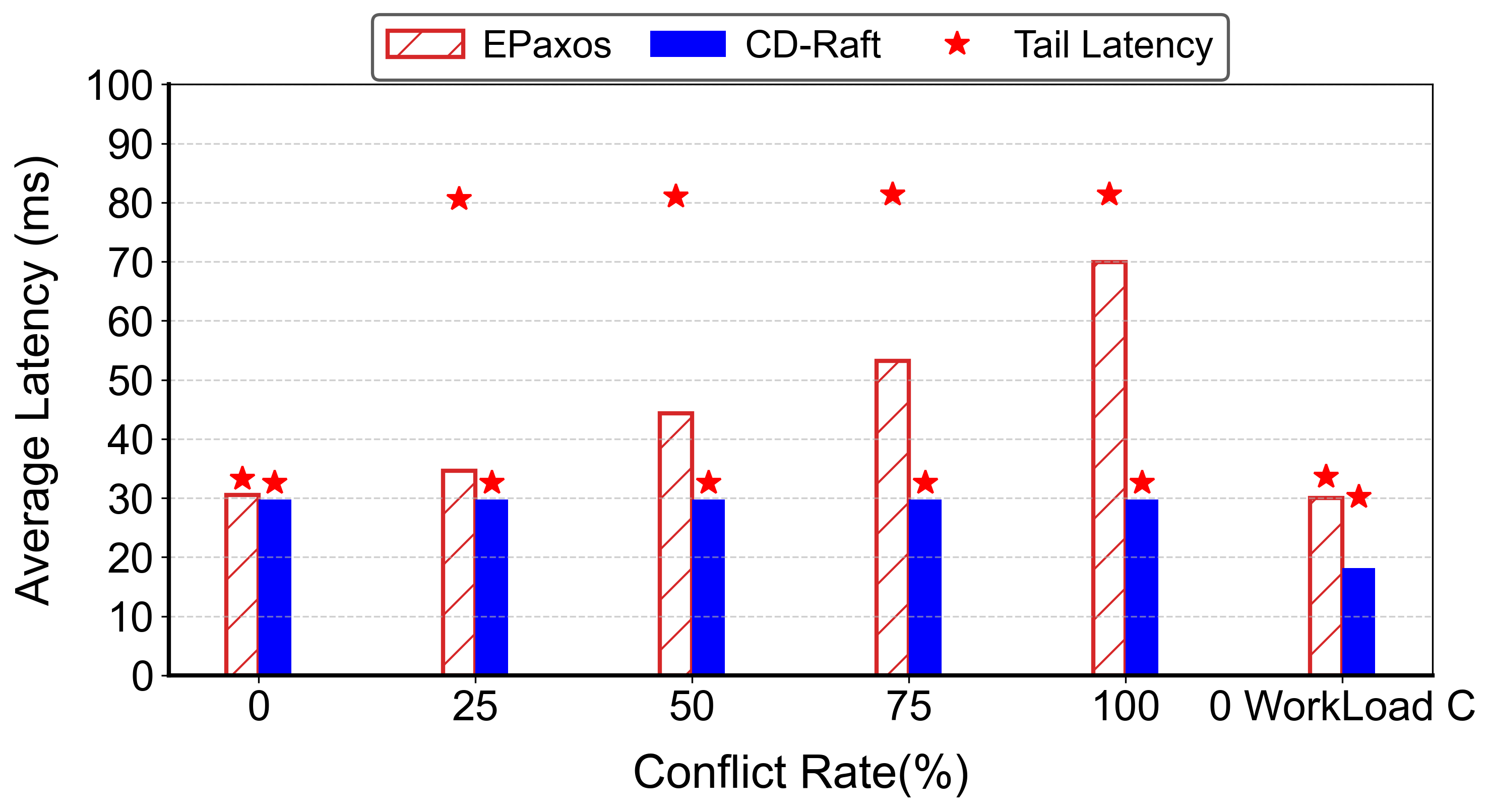}
    \caption{Comparison of latencies between EPaxos and CD-Raft under different key-value pair conflict rates.}
    \label{cpepaxos} 
    \vspace{-0.6cm}
\end{figure}
\vspace{-0.1cm}
\subsection{Impacts of Geographical Topology}
\label{geographical topology}
To comprehensively evaluate the universality and performance advantages of CD-Raft under different geographical network topologies, this experiment selected four city domains representing different regions of China: Beijing (BJ), Shanghai (SH), Guangzhou (GZ), and Guiyang (GY).  
From these, four distinct sets of three-domain cluster configurations were constructed for testing.
These configurations include: Beijing-Shanghai-Guiyang, Beijing-Shanghai-Guangzhou, Beijing-Guangzhou-Guiyang, and Shanghai-Guangzhou-Guiyang. 
Each client performed 1 million requests, evenly split between reads and writes (50\% each).
In the benchmark performance comparisons, the leaders of Raft and PigPaxos and the \textit{Global Leader} of FR used a fixed deployment: if the configuration included the Beijing domain, deployment was in Beijing; if it was the Shanghai-Guangzhou-Guiyang configuration without Beijing, deployment was in the Shanghai domain.

As Fig.\ref{num} plots, CD-Raft achieves the lowest latency in all cases.
This demonstrates that CD-Raft performs well under different geographical network topologies.
\vspace{-0.1cm}
\subsection{Impacts of Domain Count}
\label{domain count}
To evaluate the scalability of CD-Raft under varying domain scales, this experiment analyzes how protocol performance changes as the number of domains increases. 
We utilized four geographical domains to incrementally construct cluster configurations of three scales: 2-domain (Beijing-Shanghai), 3-domain (Beijing-Shanghai-Guangzhou), and 4-domain (Beijing-Shanghai-Guangzhou-Guiyang). 
Each client performed 1 million requests, evenly split between reads and writes (50\% each).

As Fig.\ref{iner} plots, CD-Raft achieves the lowest latency in all cases.
As the number of domains increases, CD-Raft's advantages become more obvious, because CD-Raft only requires two domains to commit, while Raft requires more than half of the nodes to commit. 
The system becomes unavailable only when both the \textit{Global Leader} domain and another domain fail simultaneously, requiring one domain to recover.


\vspace{-0.1cm}
\subsection{Comparison with the Leaderless Consensus Protocol}
\label{leaderless}
This experiment measures and compares the average latency of CD-Raft and the leaderless protocol EPaxos under different key-value pair conflict rates.
We set five levels for the key-value pair conflict rate: 0\% (no conflict), 25\%, 50\%, 75\%, and 100\% (full conflict).
All tests were conducted under a standard three-domain configuration.
Each client performed one million write requests.

Fig.~\ref{cpepaxos} presents the average latencies of CD-Raft and EPaxos under different key-value pair conflict rates.
The experimental results show that CD-Raft exhibits stable and excellent performance in all cases.
EPaxos's performance, however, deteriorates as the conflict rate increases.
Specifically, when there is no conflict (0\% conflict rate), both EPaxos and CD-Raft exhibit excellent performance.
As the conflict rate increases from 25\% to 100\%, the average latency reduction of CD-Raft compared to EPaxos significantly widens from 14.07\% to 57.49\%, while its tail latency is reduced by an average of 59.87\%.
These results highlight CD-Raft's excellent performance and stability in environments with data conflicts.

Furthermore, we evaluated read performance at a 0\% conflict rate (0 Workload C). 
EPaxos' read requests also need to go through the consensus process, which requires at least one cross-domain RTT.
For CD-Raft, when the client and the \textit{Global Leader} are in the same domain, they can read directly from the \textit{Global Leader} without cross-domain RTT.
Consequently, CD-Raft demonstrates a significant decrease in read latency compared to EPaxos.

\section{Related Work}
\label{sectio5}
In cross-domain environments with cross-domain disaster tolerance requirements, we have compared the following consensus protocols as shown in Table \ref{tab:consensus_protocols}.
PigPaxos \cite{pigpaxos} uses a hierarchical architecture to reduce cross-domain communication, and Mencius \cite{mencius} rotates leaders to mitigate single-point bottlenecks.
GeoLM\cite{geolm} adopts a leader management strategy that aims to improve the performance of Raft in cross-domain environments.
However, GeoLM still requires two cross-domain RTTs to handle write requests from clients outside the leader domain.
Although WPaxos\cite{WPaxos} and Flexible Paxos\cite{flexible} have different quorums, they still require two cross-domain RTTs to ensure cross-domain disaster tolerance.
In Fast Paxos\cite{fastpaxos}, the client directly sends the request to all replicas, reducing the latency by 0.5 cross-domain RTTs.
Speculative Paxos \cite{sp} and Network-Ordered Paxos \cite{sp2} rely on specialized networks to ensure the order in which requests arrive at replicas, which limits their generality.
CURP\cite{CURP} and EPaxos\cite{EPaxos} need to determine inter-request dependencies, thus lacking generality. 
Furthermore, when requests conflict, they require additional RTTs to resolve those conflicts.
Our CD-Raft requires only one cross-domain RTT, offers strong generality, and also reduces the amount of cross-domain communication.

\begin{table}[t]
\centering
\large 
\caption{Comparison of Consensus Protocols.}
\resizebox{\columnwidth}{!}{
\begin{tabular}{|c|c|c|c|}
\hline
\textbf{Protocol} & \textbf{Cross-domain RTT} & \textbf{Generality} & \textbf{Cross-domain Communication} \\ \hline
Raft              & 2            & Strong             & Many           \\ \hline
Pigpaoxs          & 2            & Strong             & Few            \\ \hline
Mencius           & 2            & Strong             & Many           \\ \hline
GeoLM             & 2            & Strong               & Many           \\ \hline
WPaxos            & 2            & Strong             & Many           \\ \hline
Flexible Paxos    & 2            & Strong             & Many           \\ \hline
Fast Paxos        & 1.5          & Strong             & Many           \\ \hline
Speculative Paxos & 1            & Weak               & Many           \\ \hline
Network-Ordered   & 1            & Weak               & Many           \\ \hline
CURP              & 1 or 2            & Weak               & Many           \\ \hline
EPaxos            & 1 or 2            & Weak               & Many           \\ \hline
CD-Raft           & 1            & Strong             & Few            \\ \hline
\end{tabular}
}
\label{tab:consensus_protocols}
\vspace{-0.5cm}
\end{table}

\section{Conclusion}
\label{sectio6}
In cross-domain environments, cross-domain RTT becomes a major performance bottleneck for distributed consensus protocols. As a commonly used consensus protocol, Raft faces the challenge of high latency when maintaining consistency in cross-domain environments, especially when clients and the leader are located in different domains. To address this issue, we propose an improved consensus protocol, CD-Raft. This is an optimized Raft protocol, which significantly reduces the high latency caused by cross-domain interactions through \textit{Fast Return} strategy and \textit{Optimal Global Leader Position} strategy. CD-Raft is efficient,
with the potential for wide application in cross-domain data center environments.

\bibliographystyle{IEEEtran}
\bibliography{ref}

\end{document}